\documentclass{ar2e}
\begin{document}
\input epsf.tex    

\input psfig.sty

\jname{Annu. Rev. Astro. Astrophys.}
\jyear{Sept 2007}
\jvol{Vol 45}
\ARinfo{}

\title{Physical Properties of Wolf-Rayet Stars}

\markboth{Paul A. Crowther}{Wolf-Rayet Stars}

\author{Paul A. Crowther \affiliation{Department of Physics \& Astronomy,
University of Sheffield, Hounsfield Road, Sheffield, S3 7RH, United Kingdom,
email: Paul.Crowther@sheffield.ac.uk}}

\begin{keywords} stars: Wolf-Rayet; stars: fundamental parameters; stars:
evolution; stars: abundances \end{keywords}

\begin{abstract} The striking broad emission line spectroscopic
appearance of Wolf-Rayet (WR) stars has long defied analysis, due to the
extreme physical conditions within their line and continuum forming regions.
Recently, model atmosphere studies have advanced sufficiently to enable
the determination of stellar temperatures, luminosities,
abundances, ionizing fluxes and wind properties. The observed
distributions of nitrogen (WN) and carbon (WC) sequence WR stars in the
Milky Way and in nearby star forming galaxies are discussed; these imply
lower limits to progenitor masses of $\sim$25, 40, 75 $M_{\odot}$ for
hydrogen-depleted (He-burning) WN, WC, and H-rich (H-burning) WN stars,
respectively. WR stars in massive star binaries permit 
studies of wind-wind interactions and dust formation in WC systems.
They also show that WR stars have typical masses of 10--25$M_{\odot}$, 
extending up to 80$M_{\odot}$  for H-rich WN stars.
Theoretical and observational evidence that WR winds depend on
metallicity is presented, with implications
for evolutionary models, ionizing fluxes, and the role of WR stars within
the context of core-collapse supernovae and long-duration gamma ray
bursts. \end{abstract}

\maketitle

\section{Introduction}

Massive stars dominate the feedback to the local interstellar medium
(ISM)  in star-forming galaxies via their stellar winds and ultimate
death as core-collapse supernovae. In particular, Wolf-Rayet (WR)  
stars typically have wind densities an order of magnitude higher than
massive O stars. They contribute to the chemical enrichment of
galaxies, they are the prime candidates for the
immediate progenitors of long, soft Gamma Ray Bursts (GRBs, Woosley \&
Bloom 2006), and they provide a signature of high-mass star formation in
galaxies (Schaerer \& Vacca 1998).

Spectroscopically, WR stars
are spectacular in appearance, with strong, broad emission lines instead
of the narrow absorption lines which are typical of normal stellar
populations (e.g. Beals 1940). The class are named after Wolf \& Rayet (1867)
who identified three stars in Cygnus with such broad emission lines. 
It was immediately apparent that their spectra came in two
flavours, subsequently identified as those with strong lines of helium
and nitrogen (WN subtypes) and those with strong helium, carbon, and
oxygen (WC and WO subtypes).  Gamov (1943) first suggested that the
anomalous composition of WR stars was the result of nuclear processed
material being visible on their surfaces, although this was not
universally established until the final decade of the 20th Century 
(Lamers et al. 1991).  
Specifically, WN and WC stars show the products of the CNO cycle
(H-burning) and the triple-$\alpha$ (He-burning), respectively. In
reality, there is a continuity of physical and chemical properties 
between O supergiants and WN subtypes.

Typically, WR stars have masses of 10--25 $M_{\odot}$, and are
descended from O-type stars. They spend $\sim$10\% of their  
$\sim$5Myr lifetime as WR stars (Meynet \& Maeder 2005). 
At Solar metallicity the minimum initial mass  for a star to become a WR star 
is $\sim$25 $M_{\odot}$. This corresponds closely to the 
Humphreys \& Davidson (1979)  limit for red supergiants (RSG), 
according to a comparison between the current  
temperature calibration of RSG and stellar models that allow  
for mass-loss  and rotation (e.g. Levesque et al. 2005). 
Consequently, some single WR stars are post-red supergiants
within a fairly  limited  mass range of probably 25--30$M_{\odot}$.
Evolution proceeds via an  
intermediate Luminous Blue Variable (LBV)  phase above 30$M_{\odot}$.  For 
close binaries, the critical mass for production of a WR star has no such 
robust lower limit, since Roche lobe overflow or common envelope evolution 
could produce a WR star instead of an extended RSG phase.

The strong, broad emission lines seen in spectra of WR stars are due to
their powerful stellar winds. The wind is sufficiently dense that an
optical depth of unity in the continuum arises in the outflowing material.
The spectral features are formed far out in the wind and are seen primarily
in emission.  The line and continuum formation regions are geometrically
extended compared to the stellar radii and their physical depths are
highly wavelength dependent. The unique spectroscopic signature of WR
stars has 
permitted their detection individually in Local Group galaxies (e.g.  
Massey \& Johnson 1998; Massey 2003), 
collectively within knots of local star forming
galaxies (e.g. Hadfield \& Crowther 2006), and as significant contributors
to the average rest-frame UV spectrum of Lyman Break Galaxies (Shapley et
al. 2003).

The present review focuses on observational properties of classical
Wolf-Rayet stars in the Milky Way and beyond, plus physical and chemical
properties determined from spectroscopic analysis, plus comparisons with
interior evolutionary models, and provides revisions to the topic
with respect to the excellent Abbott \&  Conti (1987) review. 
Low mass ($\sim0.6 M_{\odot}$) 
central stars of Planetary Nebulae displaying a Wolf-Rayet 
spectroscopic appearance (denoted [WR]) are not considered. Nevertheless,
analysis tools discussed here are common to both types of star 
(e.g. Crowther et al. 2006a).

\section{Observed Properties}

\subsection{Spectral Properties and Spectral Classification}

Visual spectral classification of WR stars is based on emission line
strengths and line ratios following Smith (1968a).
WN spectral subtypes follow a 
scheme involving line ratios of N\,{\sc iii-v} and He\,{\sc i-ii},
ranging from WN2 to WN5 for `early WN' (WNE) stars, and WN7 to WN9 for
`late WN' (WNL) stars, with WN6 stars either early or late-type.
A 'h' suffix may be used to indicate the presence of 
emission lines due to hydrogen (Smith, Shara \& Moffat 1996).

Complications arise for WN stars with intrinsically weak emission lines.
For example, WR24 (WN6ha) has a He\,{\sc ii} $\lambda$4686 emission
equivalent width that is an  
order of magnitude smaller than those in 
some other WN6 stars; the `ha' 
nomenclature indicates that hydrogen is seen both in 
absorption and emission.
From a standard spectroscopic viewpoint, such stars possess mid to late
WN spectral classifications. However, their appearance is rather more
reminiscent of Of stars than classic WN stars,
since there exists a continuity of properties between normal O
stars and late-type WN stars. These stars are widely believed to be
massive O stars with relatively strong stellar winds at a rather early
evolutionary stage. They are believed not to represent the
more mature, classic He-burning WN stars. 

Smith, Crowther \& Prinja (1994)  extended the WN sequence to very
late WN10--11 subtypes in order to include a group of emission line
stars originally classified as Ofpe/WN9 (Bohannan \& Walborn 1989). WN11
subtypes closely resemble extreme early-type B supergiants except for
the presence of He\,{\sc ii} $\lambda$4686 emission.  A quantitative
comparison of optical line strengths in Of and WNL stars is presented in
figure 8 of Bohannan \& Crowther (1999). R127 (WN11)
in the Large Magellanic Cloud (LMC) was later identified as a LBV (Stahl
et al. 1983), whilst a famous Galactic LBV, AG Car exhibited a WN11-type
spectrum at visual minimum (Walborn 1990; Smith et al. 1994).

Various multi-dimensional classification 
systems have been proposed for WN stars; they
generally involve
line strengths or widths, such that strong/broad lined stars have been
labelled WN-B (Hiltner \& Schild 1966), WN-s (Hamann, Koesterke \&
Wessolowski 1993) or WNb (Smith, Shara \& Moffat 1996). Of these, none
have generally been
adopted.  From a physical perspective, strong- and weak-lined 
WN stars do form useful sub-divisions. Therefore we shall define
weak (-w) and strong (-s) WN stars as those with He\,{\sc ii} $\lambda$5412
equivalent widths smaller than or larger than 40\,\AA. 
An obvious limitation of such an approach is
that intrinsically strong-lined WN stars  would be diluted by binary 
companions or nearby stars in spatially crowded regions 
and so might not be identified as such. 
WNE-w stars tend
to exhibit triangular line profiles rather than the more typical
Gaussian lines of WNE-s stars (Marchenko et al. 2004), since one
observes material much closer to the stellar core that is being
strongly accelerated. 

WC spectral subtypes depend on the line ratios 
of C\,{\sc iii} and C\,{\sc iv} lines along with the appearance of O\,{\sc 
iii-v}, spanning WC4 to WC9 subtypes, for which WC4--6 stars are `early' 
(WCE) and WC7--9 are `late' (WCL). Rare, oxygen-rich WO stars form an 
extension 
of the WCE sequence, exhibiting strong O\,{\sc vi} 
$\lambda\lambda$3811-34 emission (Kingsburgh,
Barlow \& Storey 1995). The most recent scheme involves WO1 to WO4 
subtypes depending on the relative strength of O\,{\sc v-vi} and C\,{\sc 
iv}  emission lines  (Crowther, De Marco \& Barlow 
1998). Finally, C\,{\sc iv} $\lambda$5801-12 appears unusually
strong in an otherwise normal WN star in a few cases, 
leading  to an intermediate WN/C classification (Conti
\& Massey 1989).  WN/C stars are indeed considered to be at an
intermediate evolutionary phase between the WN and WC stages. 

\begin{figure}
\centerline{\psfig{file=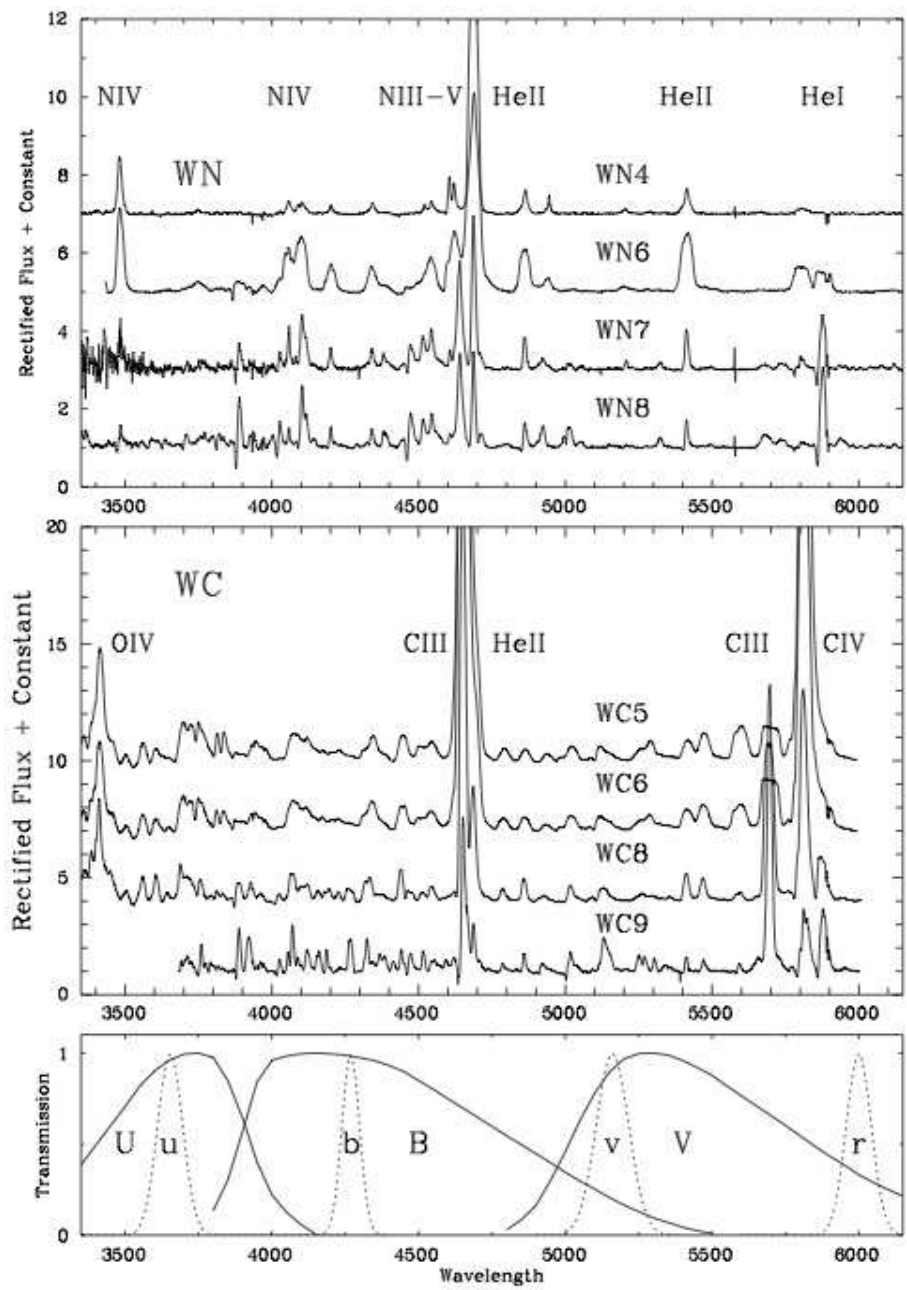,width=4.8in,angle=0}}
\caption{Montage of
optical spectroscopy of Milky Way WN and WC stars together with the 
Smith (1968b) $ubv$ and Massey (1984) $r$ narrow-band and 
Johnson UBV broad-band filters}
\label{wnc-montage}
\end{figure}

Representative examples of WN and WC stars are  presented in 
Figure~\ref{wnc-montage}. 
Various X-ray to mid-IR spectroscopic datasets of Galactic Wolf-Rayet
stars are presented in Table~\ref{atlas}, including 
extreme ultraviolet synthetic spectra from model atmospheres
(Smith, Norris \& Crowther 2002; Hamann \& Gr\"{a}fener 2004).

\subsection{Absolute magnitudes}

WR stars cannot be
distinguished from normal hot stars using UBV photometry. 
Broad-band visual measurements 
overestimate the true continuum level in extreme cases by up to 1
magnitude, or more typically 0.5 mag for single early-type WR stars due to
their strong emission-line spectra.  Consequently, Westerlund (1966)
introduced narrow-band
$ubyr$ filters that were specifically designed
to minimize the effect of WR emission lines (although their effect cannot be
entirely eliminated). These 
passbands were later refined by Smith (1968b) and by Massey
(1984), such that most photometry of WR stars has used the $ubvr$ filter
system, which is compared to Johnson UBV filters in
Fig.~\ref{wnc-montage}.

As with normal stars, $ubv$ photometry permits a determination of the 
interstellar extinction, $A_v$. Let us adopt a typical ratio of total, $A_{V}$
to selective, $E(B-V)=A_B - A_V$ extinction, $R_V = A_V/E(B-V)= 3.1$. Following
Turner (1982), the  broad-band and narrow-band optical indices for 
WR stars are then related by: 
\begin{eqnarray*}
A_v & = & 4.12 \, E_{b-v} = 3.40 \, E_{B-V} = 1.11 \, A_V
\end{eqnarray*} 
A direct determination of WR distances via stellar parallax
is only possible for $\gamma$ Vel (WC8+O) using
{\it Hipparcos}, and even that remains controversial (Millour et al. 2007). 
Otherwise, cluster or association membership is used to
provide an approximate absolute magnitude-spectral type calibration for
Milky Way WR stars. The situation is much better for WR stars in the
Magellanic Clouds, although not all subtypes are represented. Typical
absolute magnitudes range from $M_{v}$ = --3~mag at earlier subtypes to 
--6~mag
for late subtypes, or exceptionally --7~mag for hydrogen-rich WN stars.
The typical spread is $\pm$0.5 mag at  individual subtypes.

\subsection{Observed distribution}\label{evolution}

Conti (1976) first proposed that a massive O star may lose a significant
amount of mass via stellar winds, revealing first the H-burning products
at its surface, and subsequently the He-burning products.
 These evolutionary
stages are 
spectroscopically identified with the WN and WC types. 
This general picture has since become known as the `Conti scenario'.
Such stars should
be over-luminous for their mass, in accord with observations of WR stars
in binary systems. Massey (2003) provides a more general overview of 
massive stars within Local Group galaxies. 

\subsubsection{WR stars in Milky Way}

Wolf-Rayet stars are located in or close to massive star forming regions
within the Galactic disk. A catalogue is provided by van der Hucht
(2001). A quarter of the known WR stars in the Milky Way reside
within massive clusters at the Galactic centre or in Westerlund 1 (van der 
Hucht 2006).  From membership of WR stars in 
open clusters, Schild \& Maeder (1984) and Massey, DeGioia-Eastwood \&
Waterhouse (2001) investigated the initial masses of WR stars empirically.
A revised compilation is provided  in Crowther et al. (2006b). 

Overall, hydrogen-rich WN stars (WNha) are observed in young, massive 
clusters;  their main-sequence turn-off masses (based on Meynet et al. 1994 
isochrones) suggest initial  masses of $65-110 M_{\odot}$, and are
believed to be core-H burning (Langer et al. 1994;  Crowther et al. 1995a). 
Lower-mass progenitors of 40--50$M_{\odot}$ are suggested for classic mid-WN, 
late WC, and WO stars. Progenitors of some early WN stars appear to
be less massive still, suggesting an initial-mass cutoff for WR stars 
at Solar metallicity around 25$M_{\odot}$.

From an evolutionary perspective, the absence of RSGs at high luminosity
and presence of H-rich WN stars in young massive clusters suggests the
following variation of the Conti scenario in the Milky Way, i.e. for stars
initially more massive than $\sim 75 M_{\odot}$ 
 \[ O \rightarrow {\rm WN (H-rich}) \rightarrow
{\rm LBV} \rightarrow {\rm WN (H-poor)} \rightarrow {\rm WC} \rightarrow
{\rm SN\,Ic}, \] whereas for stars of initial mass from $\sim 40-75
M_{\odot}$, \[ O \rightarrow {\rm LBV} \rightarrow {\rm WN (H-poor)}
\rightarrow {\rm WC} \rightarrow {\rm SN\,Ic}, \] and for stars of initial
mass in the range 25--40$M_{\odot}$, \[ O \rightarrow {\rm LBV/RSG} 
\rightarrow
{\rm WN (H-poor)} \rightarrow {\rm SN\,Ib}.\] 
Indeed, the role of the LBV phase is
not yet settled -- it may be circumvented entirely in some 
cases; it may  follow the RSG stage, 
or it may even dominate pre-WR mass-loss for the most massive stars
(Langer et al. 1994; Smith \& Owocki 2006). Conversely, the presence of 
dense, circumstellar
shells around Type~IIn SN indicates that some massive stars may even
undergo core-collapse during the LBV phase (Smith et al. 2007).
Remarkably few Milky Way clusters host both RSG and WR stars,
with the  notable exception of Westerlund~1 (Clark et al. 2005);
this suggests that the mass range common to both populations is fairly narrow.

Although optical narrow-band surveys (see below) have proved very successful 
for  identifying WR stars in the Solar neighbourhood, 
only a few hundred WR stars are known in the Milky Way, whilst many
thousands are expected within the Galactic disk (van der Hucht 2001).  
Consequently, near-IR narrow-band imaging surveys together with 
spectroscopic follow-up  may be considered for more
extensive surveys to circumvent high interstellar extinction (Homeier et al.  
2003). Limitations of IR emission-line surveys are that fluxes of
near-IR lines  are much  weaker than those of optical lines, 
Also, no strong WR lines are  common to 
all spectral types in  the frequently used K band. 
An added complication is 
that some WC stars form dust which heavily dilutes emission line fluxes 
longward of the visual. Nevertheless, infrared surveys are presently underway 
to get an improved census of WR stars in the Milky Way. Alternatively, WR 
candidates may be identified from their near- to  mid-IR colours, which, as 
in other early-type supergiants, are  unusual due to strong 
free-free excess emission (Hadfield et al. 2007).

\subsubsection{WR stars in the Local Group}

\begin{figure}
\centerline{\psfig{file=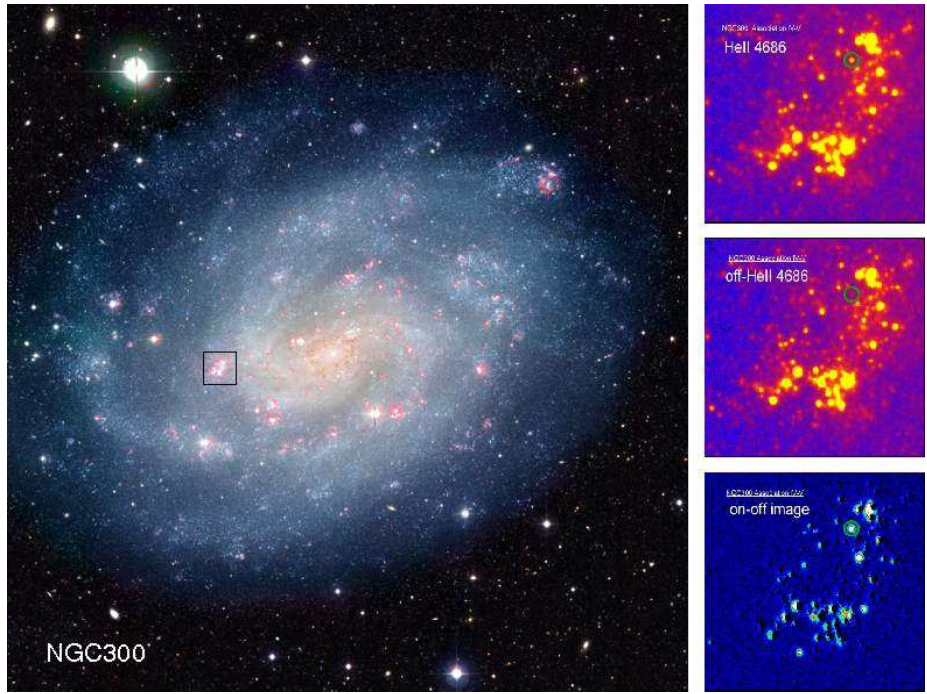,width=4.8in,angle=0}}
\caption{Composite ESO Wide Field Imager B, V, R and H$\alpha$ image of 
NGC~300 obtained at the MPG/ESO 2.2m telescope (Press Photo 18a-h/02).
A box marks the OB association IV-V (40$\times40$ arcsec).
Narrow-band images of the association are shown on the right,
centred at $\lambda$4684 (He\,{\sc ii} 
4686,  top) and $\lambda$4781 (off-He\,{\sc ii} 4686, middle), plus a 
difference  image (on-off, bottom) 
obtained with ESO VLT/FORS2 (Schild et al. 2003).
A number of WR stars showing a He\,{\sc ii} excess (white) can be seen in
the lower right image, 
including a WC4 star (green circle in all FORS2 images).}
\label{ngc300}
\end{figure}

WR stars have typically been discovered via
techniques sensitive to their unusually broad emission-line spectra, based
on objective prism searches or interference filter imaging (see
Massey 2003).  Narrow-band
interference filter techniques have been developed (e.g. Moffat, Seggewiss
\& Shara 1985; Massey, Armandroff \& Conti 1986) that distinguish strong
WR emission lines at He\,{\sc ii} $\lambda$4686 (WN stars) and C\,{\sc
iii} $\lambda$4650 (WC stars) from the nearby continuum. Such techniques have 
been applied to regions of the Milky Way disk, the Magellanic Clouds and other
nearby galaxies.  An example of this approach is presented in
Figure~\ref{ngc300} for the spiral galaxy NGC~300 ($d \sim$ 2~Mpc). A
wide-field image of NGC~300 is presented, with OB complex IV-V indicated,
together with narrow-band images centred at $\lambda$4684 (He\,{\sc ii}
4686) and $\lambda$4781 (continuum).  Several WR stars are seen in
the difference (He\,{\sc ii}-continuum)  image, including an apparently
single WC4 star (Schild et al. 2003).

It is well established that the absolute number of WR stars and their
subtype distribution are metallicity dependent. N(WR)/N(O)$\sim$0.15 
in the relatively metal-rich Solar Neighbourhood (Conti et al.  
1983; van der Hucht 2001), yet N(WR)/N(O)$\sim$0.01 in the 
metal-deficient SMC on the basis of only 12 WR stars (Massey, Olsen \& 
Parker 2003) versus $\sim$1000 O stars (Evans et al. 2004). It is 
believed that the 
majority of Galactic WR stars are the result of single-star evolution, 
yet some stars (e.g. V444 Cyg) result from close binary evolution 
(Vanbeveren et al. 1998).

Similar relative numbers of WN to WC stars are observed 
in the Solar Neighbourhood
(Hadfield et al. 2007). In contrast,
WN stars exceed WC stars by a factor of
$\sim$5 and $\sim$10 for the LMC and SMC, respectively (Breysacher,
Azzopardi \& Testor 1999; Massey, Olsen \& Parker 2001).  
At low metallicity the reduced
WR population and  the relative dominance of WN subtypes most likely result
from the metallicity dependence of winds from their evolutionary
precursors (Mokiem et al. 2007). Consequently,
only the most massive single stars reach the WR phase in metal-poor
environments. 
Single stars reaching the WC phase at high metallicity may end
their lives as a RSG or WN stars in a lower metallicity environment. 
As such,  one might suspect that most WR stars at low metallicity 
are formed via binary evolution. However, 
Foellmi, Moffat \& Guerrero (2003a) suggest a similar WR binary
fraction for the SMC and Milky Way.

Not all WR subtypes are observed in all environments.  Early WN and WC
subtypes are preferred in metal-poor galaxies, such as the SMC (Massey et al.
2003), while late WC stars are more common at super-Solar metallicities, such
as M83 (Hadfield et al. 2005) 
Line widths of early WC and WO stars are higher
than late WC stars, although width alone is not a defining criterion for
each spectral type. The correlation between WC subclass and line width
is nevertheless strong (Torres, Conti \& Massey 1986).
The subtype distributions of WR stars in the Solar
Neighbourhood, LMC, and SMC are presented in Figure~\ref{wrpop}. We shall
address this aspect in Sect~\ref{metallicity}.

\subsubsection{WR galaxies}

Individual WR stars may, in general, be resolved in Local Group galaxies
from ground-based observations,
whilst the likelihood of contamination by nearby sources increases at
larger distances.  For example, a typical slit width of 1$''$ at the 2\,Mpc
distance of NGC~300 corresponds to a spatial scale of $\sim$10~pc.
Relatively isolated WR stars have been identified, albeit in the
minority (recall Figure~\ref{ngc300}). This is even
more problematic for more distant galaxies such as M~83 where the great
majority of WR stars are observed in clusters or associations (Hadfield et
al. 2005). So-called `WR galaxies' are typically starburst regions
exhibiting spectral features from tens, hundreds, or even thousands of WR
stars (Schaerer, Contini \& Pindao 1999).

Average Milky Way/LMC WN or WC line fluxes (Schaerer \& Vacca 1998) are
typically used to calculate stellar populations in WR galaxies. These should be
valid provided that the line fluxes of WR templates do not vary with
environment. However, it is well known that SMC WN stars possess weak emission
lines (Conti, Garmany \& Massey 1989). In spite of small statistics and a
large scatter, the mean He\,{\sc ii} $\lambda$4686 line luminosity of
WN2--4 stars in the LMC is 10$^{35.9}$ erg\,s$^{-1}$, a factor of five times
higher than the mean of equivalent stars in the SMC (Crowther \&
Hadfield 2006).  The signature of WN stars is most 
readily seen in star forming galaxies at He\,{\sc ii} $\lambda$1640, 
where the dilution from other stellar types is at its weakest 
(e.g. Hadfield \& Crowther 2006).  
The  strongest  UV, optical,  and near-IR lines indicate flux ratios of 
$I$(He\,{\sc ii}  1640)/$I$(He\,{\sc ii} 4686)$\sim$10 and $I$(He\,{\sc 
ii}  4686)/$I$(He\,{\sc ii} 1.012$\mu$m)$\sim$6  for WN stars spanning SMC 
to Milky Way metallicities. 

Similar comparisons for WC stars are hindered because the only 
carbon-sequence WR stars at the low metallicity of the SMC and IC~1613 are
WO stars. Their emission line fluxes 
are systematically weaker than WC stars in the LMC and
Milky Way (Kingsburgh et al. 1995; Kingsburgh \& Barlow 1995; Schaerer \& 
Vacca 1998). The mean C\,{\sc iv} $\lambda\lambda$5801-2 
line luminosity of WC4 stars in the LMC is $10^{36.5}$ erg\,s$^{-1}$ 
(Crowther \&  Hadfield 2006). Again, detection of WC stars is favoured
via ultraviolet spectroscopy of C\,{\sc iv} $\lambda$1550. For WC stars,
the strongest UV, optical, and near-IR lines possess flux ratios of
$I$(C\,{\sc iv}  1548-51)/$I$(C\,{\sc iv} 5801-12)$\sim$6 and $I$(C\,{\sc 
iv}  5801-12)/$I$(C\,{\sc iv} 2.08$\mu$m)$\sim$15.

\subsection{Binary statistics and masses}

The observed binary fraction amongst  Milky Way WR stars is 40\% (van
der Hucht 2001), either from spectroscopic or indirect techniques. Within 
the low metallicity Magellanic Clouds, close binary evolution  would be 
anticipated to play a greater role because of the diminished role of
O star mass-loss in producing single WR stars. 
However, where detailed studies have 
been carried out (Bartzakos, Moffat \& Niemela 2001; Foellmi, Moffat \& 
Guerrero 2003ab), a similar binary fraction to the Milky Way has been 
obtained (recall Figure~\ref{wrpop}), so metallicity-independent LBV
eruptions may play a dominant role.

The most robust method of measuring stellar masses is from Kepler's third
law of motion, particularly for eclipsing double-lined (SB2)  systems,
from which the inclination may be derived.  Orbital inclinations may also
be derived from linear polarization studies (e.g. St-Louis et al. 1993) or
atmospheric eclipses (Lamontagne et al. 1996).
Masses for Galactic WR stars are included in the van der Hucht
(2001) compilation, a subset of which are presented in
Figure~\ref{binary_masses} together with some more recent results. 
WC masses  span a narrow range of
9--16$M_{\odot}$, whilst WN stars span a very wide range of
$\sim$10--83$M_{\odot}$, and in some cases exceed their OB companion, i.e.  
$q = M_{\rm WR}/M_{O} > 1$ (e.g. WR22: Schweickhardt et al. 1999). WR20a
(SMSP2) currently sets the record for the highest orbital-derived
mass of any star, with $\sim 83 M_{\odot}$ for each WN6ha component (Rauw
et al. 2005). As discussed above, such stars are H-rich, extreme O stars
with strong winds rather than classical H-poor WN stars. They are 
a factor 
of two lower in mass 
than the apparent $\sim150 M_{\odot}$ stellar mass limit (Figer 
2005), such that still more extreme cases may await discovery. 
Spectroscopic measurement of masses via surface gravities using 
photospheric lines is not possible for WR stars due to their dense stellar winds.

\subsection{Rotation velocities}

Rotation is very difficult to measure in WR stars, since
photospheric features -- used to estimate $v \sin i$ in normal stars --
are absent.  Velocities of 200--500 km\,s$^{-1}$ have been
inferred for WR138 (Massey 1980) and WR3 (Massey \& Conti 1981), although
these are not believed to represent rotation velocities, since the
former has a late-O binary companion, and the absorption lines of the
latter are formed within the stellar wind (Marchenko et al. 2004).
Fortunately, certain WR stars do harbour large scale structures, from which
a rotation period may be inferred (St-Louis et al. 2007).

Alternatively, if WR stars were rapid rotators, one would expect strong 
deviations from spherical symmetry due to gravity darkening (Von Zeipel 
1924; Owocki, Cranmer \& Gayley 1996). Harries, 
Hillier \& Howarth (1998) studied linear spectropolarimetric datasets for 29 
Galactic WR stars, from which just four single WN stars
plus one WC+O binary revealed a strong line effect, suggesting significant
departures from spherical symmetry. They presented radiative transfer
calculations which suggest that the observed continuum polarizations for
these stars can be matched by models with equator to pole density ratios
of 2--3.  Of course, the majority of Milky Way WR stars do not show a
strong linear polarization line effect (e.g. Kurosawa, Hillier \&
Schulte-Ladbeck (1999).

\subsection{Stellar wind bubbles}

Ring nebulae are observed for a subset of WR stars. These are believed to
represent material ejected during the RSG or LBV phases that is 
photo-ionized by the WR star. The first known
examples, NGC~2359 and NGC~6888, display a shell morphology, although many
subsequently detected in the Milky Way and Magellanic Clouds exhibit a
variety of spatial morphologies (Chu, Treffers \& Kwitter 1983; Dopita et
al. 1994). Nebulae are predominantly associated with young WR stars i.e.
primarily WN subtypes, with typical electron densities of 10$^{2}$ 
cm$^{-3}$
to  10$^{3}$ cm$^{-3}$ (Esteban et al. 1993).

Ring nebulae provide information on evolutionary links between 
WR stars and their precursors (Weaver et al. 1977).
Once a massive star has reached the WR phase, its fast wind will
sweep up the material ejected during the immediate precursor 
(LBV or RSG)  slow wind. The dynamical evolution of gas around 
WR stars with such
progenitors has been discussed 
by Garc\'{i}a-Segura, MacLow \& Langer (1996a)  
and Garc\'{i}a-Segura, Langer \& MacLow (1996b). 
Esteban et al. (1993) attempted
to derive WR properties indirectly from H\,{\sc ii} regions associated
with selected Milky Way stars (see also Crowther et al. 1999).
Unfortunately, relatively few H\,{\sc ii} regions are associated with
individual WR stars, and for the majority of these, the nebular parameters are
insufficiently well constrained to distinguish between different stellar
atmosphere models.

\section{Physical Parameters}\label{atmos}

\subsection{Radiative Transfer}

Our interpretation of hot, luminous stars via radiative transfer codes is
hindered with respect to normal stars by several effects. First, the
routine assumption of LTE breaks down for high-temperature stars. In
non-LTE, the determination of populations uses rates which are functions
of the radiation field, itself a function of the populations.
Consequently, it is necessary to solve for the radiation field and
populations iteratively.
Second, the problem of accounting for the effect of millions of
spectral lines upon the emergent atmospheric structure and emergent
spectrum -- known as line blanketing -- remains challenging for stars in
which spherical, rather than plane-parallel, geometry must be assumed due 
to
stellar winds, since the scale height of their atmospheres is not
negligible with respect to their stellar radii. The combination of
non-LTE, line blanketing (and availability of atomic data thereof), and 
spherical geometry has prevented the routine analysis of such stars until 
recently.

Radiative transfer is either solved in the co-moving frame, as applied by
CMFGEN (Hillier \& Miller 1998) and PoWR (Gr\"{a}fener, Koesterke \&
Hamann 2002) or via the Sobolev approximation, as used by ISA-wind (de
Koter, Schmutz \& Lamers 1993). The incorporation of line blanketing
necessitates one of several approximations. Either a `super-level'
approach is followed, in which spectral lines of a given ion are grouped
together in the solution of the rate equations (Anderson 1989), or
alternatively, a Monte Carlo approach is followed, which uses 
approximate level populations (Abbott \& Lucy 1985).

\subsection{Stellar Temperatures and Radii}

Stellar temperatures for WR stars are difficult to characterize, 
because their geometric extension is comparable with their stellar radii. 
Atmospheric models for WR stars are typically parameterized by the radius 
of the inner boundary $R_{\ast}$ at high Rosseland optical depth 
$\tau_{\rm Ross} (\sim 10$). However, only the optically thin part of the 
atmosphere is seen by the observer. The measurement of $R_{\ast}$ depends 
upon the assumption that the same velocity law holds for the visible 
(optically thin) and the invisible (optically thick) part of the 
atmosphere. 

The optical continuum radiation originates from a 
`photosphere' where $\tau_{\rm Ross} \sim 2/3$. Typical WN and WC winds 
have reached a significant fraction of their terminal velocity before they 
become optically thin in the continuum. $R_{2/3}$, the radius at 
$\tau_{\rm Ross} = 2/3$ lies at highly supersonic velocities, well beyond 
the hydrostatic domain. For example, Crowther et al. (2006a) obtain 
$R_{\ast}$ = 2.9\,$R_{\odot}$ and $R_{2/3}$ = 7.7\,$R_{\odot}$ for 
HD~50896 (WN4b), corresponding to $T_{\ast}$ = 85\,kK and $T_{2/3}$ = 
52\,kK, respectively. In some weak-lined, early-type WN stars, this is not 
strictly true since their spherical extinction is modest, in which case 
R$_{\ast} \sim R_{2/3}$ (e.g. HD~9974, Marchenko et al. 2004).

Stellar temperatures of WR stars are derived from lines from adjacent 
ionization stages of helium or nitrogen for WN stars (Hillier 1987, 1988), 
or lines of carbon for WC stars (Hillier 1989).  High stellar wind 
densities require the simultaneous determination of mass-loss rate and 
stellar temperature from non-LTE model atmospheres, since their 
atmospheres are so highly stratified. Metals such as C, N and O provide 
efficient coolants, such that the outer wind electron temperature is 
typically 8\,kK to 12\,kK (Hillier 1989).
Figure~\ref{WR_ross} compares $R_{\ast}$, $R_{2/3}$, and 
the principal optical wind line-forming region ($\log n_{e} = 10^{11}$ 
to 10$^{12}$ cm$^{-3}$) for HD~66811 ($\zeta$ Pup, O4\,I(n)f), HD~96548 
(WR40, WN8) and HD~164270 (WR103, WC9) on the same physical scale. Some 
high-ionization spectral lines (e.g. N\,{\sc v} and C\,{\sc iv} lines in 
WN8 and WC9 stars, respectively) are formed at higher densities of 
$n_{e} \geq 10^{12}$ cm$^{-3}$ in the WR winds.

Derived stellar temperatures depend sensitively upon the detailed 
inclusion of line-blanketing by iron peak elements. Inferred bolometric 
corrections and stellar luminosities also depend upon detailed metal 
line-blanketing (Schmutz 1997; Hillier \& Miller 1999). Until recently, 
the number of stars studied with non-LTE, clumped, metal line-blanketed 
models has been embarrassingly small, due to the need for detailed, 
tailored analysis of individual stars using a large number of free 
parameters. Hamann, Gr\"{a}fener \& Liermann (2006) have applied 
their grid of line-blanketed WR models to the analysis of most Galactic WN 
stars, for the most part resolving previous discrepancies between 
alternate line diagnostics, which were 
first identified by Crowther et al.  (1995b). 
To date, only a limited number of WC stars in the Milky Way and Magellanic 
Clouds have been studied in detail (e.g. Dessart et al. 2000; Crowther et 
al. 2002; Barniske, Hamann \& Gr\"{a}fener 2007). Results for
Galactic and LMC WR stars are presented in Table~\ref{WRtemp}. These range 
from 30\,kK amongst WN10 subtypes to 40\,kK at WN8 and approach 100\,kK 
for early-type WN stars. Spectroscopic temperatures are rather higher 
for WC stars, i.e. 50\,kK for WC9 stars, increasing to 70\,kK at WC8 and 
$\geq$100\,kK for early WC and WO stars

Stellar structure models predict radii $R_{\rm evol}$ that are 
significantly smaller than those derived from atmospheric models. For
example, $R_{\ast} = 2.7\,R_{\odot}$ for HD~191765 (WR134, WN6b) 
in Table~\ref{WRtemp}, versus
$R_{\rm evol} = 0.8\,R_{\odot}$ which follows from 
hydrostatic evolutionary models, namely
\begin{equation}\label{radii}
\log \frac{R_{\rm evol}}{R_{\odot}} = -1.845 + 0.338 \log 
\frac{L}{L_{\odot}}
\end{equation}
for hydrogen-free WR stars (Schaerer \& Maeder 1992).
Theoretical corrections to such radii are frequently applied,
although these are based upon fairly arbitrary assumptions which 
relate particularly to the velocity law. Consequently, a direct comparison 
between temperatures of most WR stars from evolutionary calculations and 
empirical atmospheric models is not straightforward, except that one requires 
$R_{2/3} > R_{\rm evol}$, with the difference attributed to the 
extension of the supersonic region. Petrovic, Pols \& 
Langer (2006) established that the hydrostatic cores of metal-rich WR stars 
above $\sim 15~M_{\odot}$ exceed $R_{\rm evol}$ in Eqn~\ref{radii} by 
significant factors 
if mass-loss is neglected, due to their proximity to the Eddington limit, 
$\Gamma_{e}$ = 1. Here, the Eddington parameter, $\Gamma_{e}$, is the 
ratio of  radiative acceleration due to Thompson (electron)
scattering to surface gravity and may be written as
\begin{equation}
\Gamma_e = 10^{-4.5} q_{e} \frac{L/L_{\odot}}{M/M_{\odot}}
\end{equation}
where the number of free electrons per atomic mass unit is $q_e$. 
In reality, high empirical WR mass-loss rates imply that inflated radii
are not expected, such 
that  the discrepancy in hydrostatic radii between stellar structure and 
atmospheric models has not yet been resolved.

\subsection{Stellar Luminosities}

Absolute visual magnitudes of WR stars are obtained primarily from 
calibrations obtained from cluster or association membership (van der 
Hucht 2001). Inferred bolometric corrections range from $M_{\rm bol} - 
M_{v}$ = --2.7~mag amongst very late WN stars (Crowther \& Smith 1997) to 
approximately --6 mag for weak-lined, early-type WN stars and WO stars 
(Crowther et al. 1995b; Crowther et al. 2000). Stellar luminosities of 
Milky Way WN stars range from 200,000 $L_{\odot}$ in early-type stars to 
500,000 $L_{\odot}$ in late-type stars. Hydrogen-burning O stars with 
strong stellar winds, spectroscopically identified as WNha stars, have 
luminosities in excess of $10^{6}~L_{\odot}$. For Milky Way WC stars, 
inferred stellar luminosities are $\sim$150,000 $L_{\odot}$, increasing by 
a factor of two for LMC WC stars (Table~\ref{WRtemp}).

Systematically higher spectroscopic luminosities have recently been
determined by Hamann, Gr\"{a}fener \& Liermann (2006) for Galactic WN
stars, since they adopt uniformly high $M_{v} = -7.2$~mag for all 
non-cluster member WN6--9h stars. Absolute magnitudes for normal late-type 
WN stars are subject to large uncertainties since such stars positively
shy away from clusters. As a consequence, their results suggest a bi-modal 
distribution around $300,000~L_{\odot}$ for early WN stars, and 
1--2$\times 10^{6}~L_{\odot}$ for all late WN stars. 

From stellar structure theory, there is a mass-luminosity relation for H-free 
WR stars which is described by
\begin{equation} 
\log \frac{L}{L_{\odot}} = 3.032 + 2.695 \log
\frac{M}{M_{\odot}} - 0.461 \left( \log \frac{M}{M_{\odot}} \right)^{2}.
\end{equation} 
This expression is effectively independent of the chemical composition 
since the continuum opacity is purely electron scattering (Schaerer \& 
Maeder 1992). Spectroscopic luminosities need to be corrected for the 
luminosity that powers the stellar wind, $\frac{1}{2} \dot{M} 
v_{\infty}^{2}$, in order to determine the underlying nuclear luminosity, 
$L_{\rm nuc}$. In most cases, the recent reduction in estimates of 
mass-loss rates due to wind clumping (see Sect.~\ref{clump}), plus the 
increase in derived luminosities due to metal line-blanketing indicate a 
fairly modest corrective factor.  From Table~\ref{WRtemp}, one expects 
typical masses of 10--15~$M_{\odot}$ for hydrogen-free WR stars, which 
agree fairly well with binary mass estimates (recall 
Fig.~\ref{binary_masses}). Indeed, spectroscopically derived WR masses 
obtained using this relationship agree well with binary derived masses 
(e.g. $\gamma$ Vel: De Marco et al. 2000).

\subsection{Ionizing fluxes}
  
Lyman continuum ionizing fluxes, N(LyC), are typical of mid-O stars in 
general (Table~\ref{WRtemp}). As such, the low number of WR stars
with respect to O stars would suggest that
Wolf-Rayet stars play only a minor role in the Lyman continuum
ionization budget of young star-forming regions. 
H-rich late-type WN stars provide a notable exception,
since their ionizing output compares closely to O2 stars (Walborn et al.
2004). Crowther \& Dessart (1998) showed that the WN6ha stars in 
NGC~3603 provided $\sim$20\% of the Lyman continuum ionizing photons, 
based upon calibrations of non-blanketed models for O and WR stars.

Since WR stars represent an extension of O stars to higher temperatures,
significant He\,{\sc i} continuum photons are emitted, plus strong
He\,{\sc ii} continua for a few high-temperature, low-density cases.
The primary effect of metal-line 
blanketing is to redistribute extreme UV flux to
longer wavelengths, reducing the ionization balance in the wind,
such that higher temperatures and luminosities are required to match the
observed WR emission line profile diagnostics relative to unblanketed
models. Recent revisions to estimated 
temperatures and luminosities of O stars (as
measured from photospheric lines) have acted in the reverse sense, 
relative to previous plane-parallel unblanketed model analysis, 
due to backwarming effects, as shown
in a number of recent papers (e.g. Martins, Schaerer \& Hillier 
2002; Repolust, Puls \& Herrero 2004). Common techniques are 
generally now employed for O and WR studies, such that a factor of two 
increase in N(LyC) for WR stars --  plus the reverse for O stars --
suggests that in such cases WR stars might provide close to
half of the total ionizing photons in the youngest starbursts, such
as NGC~3603.

The strength of WR winds affects the hardness of their ionizing
radiation. Atmospheric models for WR stars with dense winds produce
relatively soft ionizing flux distributions, in which extreme UV photons are
redistributed to longer wavelength by the opaque stellar wind (Schmutz,
Leitherer \& Gruenwald 1992). In contrast, for the low wind-density case, a
hard ionizing flux distribution is predicted, in which extreme UV photons pass
through the relatively transparent wind unimpeded. Consequently, the shape of
the ionizing flux distribution of WR stars depends on both the wind
density and the stellar temperature.  

We shall show in Section~\ref{mdot} that
low-metallicity WR stars possess weaker winds. In Figure~\ref{wn_grid}, we
compare the predicted Lyman continuum ionizing flux distribution from four
100\,kK WN models, in which only the low metallicity, low mass-loss rate
models predicts a prodigious number of 
photons below the He$^{+}$ edge at 228\AA. Consequently,
one expects evidence of hard ionizing radiation from WR stars (e.g. nebular
He\,{\sc ii} $\lambda$4686) solely at low metallicities. This is generally borne out
by observations of H\,{\sc ii} regions associated with WR stars (Garnett et
al. 1991). Previous studies of metal-rich regions have claimed a low limit to
the stellar mass function from indirect H\,{\sc ii} region studies at high
metallicity in which WR stars were spectroscopically detected. Gonzalez
Delgado et al. (2002) were able to reconcile a high stellar mass limit from UV
spectral synthesis techniques with a soft ionizing spectrum for the metal-rich
WR galaxy NGC~3049 by applying the Smith, Norris \& Crowther (2002) line
blanketed grid of WR models at high metallicity.

\subsection{Elemental abundances}\label{abundances}

For WR stars, it has long been suspected that abundances represented the
products of core nucleosynthesis, although it has taken the development of
non-LTE model atmospheres for these to have been empirically 
supported.

\subsubsection{WN and WN/C stars}

Balmer-Pickering decrement studies by Conti, Leep \& Perry 
(1983) concluded that
hydrogen was severely depleted in WR stars. A clear subtype effect
regarding the hydrogen content of Galactic WN stars is observed, with 
late-type 
WN
stars generally showing some hydrogen (typically $X_{H} \sim 15\pm$10\%), 
and early-type WN stars being
hydrogen-free,
although exceptions do exist. This trend breaks down within the 
lower metallicity environment of the Magellanic Clouds, notably 
the SMC (Foellmi et al 2003a). Milky Way late-type  WN stars with weak 
emission lines --  denoted as `ha' due to intrinsic absorption
lines plus the presence of hydrogen -- are universally H-rich with
$X_{H}\sim$50\% (Crowther et al. 1995a; Crowther \& Dessart 1998). 

Non-LTE analyses confirm that WN abundance
patterns are consistent with material processed by the CNO 
cycle in which these elements are used as catalysts, i.e.
\[
 ^{12}C(p,\gamma) ^{13}N (e^{-}, \nu_e) ^{13}C (p,\gamma) ^{14}N 
(p,\gamma) ^{15}O (e^{-}, \nu_e) ^{15}N (p, \alpha) ^{12}C
\]
in which $X_{N}\sim$1\% by mass is observed in Milky Way
WN stars. Carbon is highly depleted, 
with
typically $X_{C} \sim$ 0.05\%. Oxygen suffers from fewer readily 
accessible line diagnostics, but probably exhibits a similarly low mass 
fraction as carbon
(e.g. Crowther, Smith \& Hillier 1995b; Herald, Hillier \& Schulte-Ladbeck 
2001). Non-LTE analysis of transition WN/C stars reveals elemental abundances
(e.g. $X_{C} \sim$ 5\%, $X_{N} \sim$ 1\% by mass) that are in good
agreement with the hypothesis that these stars are in 
a brief transition stage between 
WN and WC (Langer 1991; Crowther, Smith \& Willis 1995c).

\subsubsection{WC and WO stars}

Neither hydrogen nor nitrogen are detected in the spectra of WC stars.
Recombination line studies using theoretical coefficients for different 
transitions  are most readily applicable to WC stars, since they show a large 
number of  lines in their optical spectra. Atomic data are most reliable for 
hydrogenic ions, such as C\,{\sc iv} and O\,{\sc vi}, so early-type WC and WO 
stars can be studied most readily. Smith \& Hummer (1988) suggested a trend
of increasing C/He from late to early WC stars, with C/He=0.04--0.7 by number
(10\% $\leq X_{C} \leq$ 60\%), revealing the products of core He burning
\[
  ^{4}He (2\alpha,\gamma)^{12}C \hspace*{2cm} {\rm and} \hspace*{2cm}  ^{12}C 
(\alpha,\gamma) ^{16}O 
\]
although significant uncertainties remain in the rate of the 
latter nuclear reaction. 
These reactions compete during helium  burning  to determine the ratio of 
carbon to oxygen at the onset of carbon burning.

In reality, optical depth effects come into play, so detailed abundance 
determinations for all subtypes require non-LTE model 
atmosphere analyses. Koesterke \& Hamann (1995) indicated refined values of 
C/He=0.1--0.5 by number (20\% $\leq X_{C} \leq$ 55\%), with
no WC subtype dependence, such that spectral types are not dictated by 
carbon abundance, contrary to suggestions by Smith \& Maeder (1991).
Indeed, LMC WC4 stars possess similar surface abundances to Milky Way WC
stars (Crowther et al. 2002), for which the He\,{\sc ii} $\lambda$5412 and C\,{\sc 
iv} $\lambda$5471 optical lines represent the primary diagnostics
(Hillier 1989). These recombination lines are formed at high densities 
of 10$^{11}$ to 10$^{12}$ cm$^{-3}$ at radii of 3--30 $R_{\ast}$ 
(recall Figure~\ref{WR_ross}).  Oxygen diagnostics in WC stars lie in the 
near-UV, such that derived oxygen abundances are rather unreliable unless 
space-based spectroscopy is available. Where they 
have been derived, one finds $X_{O} \sim$ 5--10\% for WC stars 
(e.g. Crowther et al. 2002).

Core He burning in massive stars also has the effect of transforming
$^{14}$N (produced in the CNO cycle) to neon and magnesium  via
\[
 ^{14}N (\alpha, \gamma) ^{18}F (e^{-},\nu_e) ^{18}O (\alpha,\gamma)
^{22}Ne (\alpha, n) ^{25}Mg 
 \]
and serves as the main neutron source for the s-process in massive stars.
Neon lines are extremely weak in the UV/optical spectrum of WC stars
(Crowther et al. 2002), but ground-state fine-structure lines at 
[Ne\,{\sc ii}] 12.8$\mu$m and [Ne\,{\sc iii}] 
15.5$\mu$m may  be observed via mid-IR 
spectroscopy, as illustrated for $\gamma$ Vel in van der Hucht et al. (1996).
Fine-structure wind lines  are formed at hundreds of stellar radii since 
their critical densities are of order 10$^{5}$ cm$^{-3}$.
Barlow, Roche \& Aitken (1988) came to the
conclusion that neon was not greatly enhanced in $\gamma$ Vel
with respect to the Solar case ($\sim$0.1\% by mass primarily in the
form of $^{20}$Ne) in $\gamma$ Vel
from their analysis of fine-structure lines. This was
a surprising result, since the above reaction is expected to produce
$\sim$2\% by mass of $^{22}$Ne at Solar metallicity.

Once the clumped nature of WR winds 
is taken into consideration, neon is found to be enhanced in $\gamma$ 
Vel and other WC stars (e.g. Dessart et al. 2000). The inferred neon
mass fraction is $\sim$1\% (see also Crowther, Morris \& Smith 2006a). 
Meynet \& Maeder (2003) note that the $^{22}$Ne enrichment depends 
upon nuclear reaction rates rather than stellar models, so the remaining
disagreement may suggest a problem with the relevant reaction rates.
More likely, a lower metal content is inferred from the neon abundance
than Solar metallicity evolutionary models (Z=0.020). Indeed, if 
the Solar oxygen abundance from Apslund et al. (2004) is taken into account, 
a revised metal content of Z=0.012 for the Sun is impled. Allowance for 
depletion of heavy elements due to diffusion in the 4.5~Gyr old Sun suggests a 
Solar neighbourhood metallicity of Z=0.014 (Meynet, private communication).
It is likely that allowance for a reduced CNO content would bring 
predicted and measured Ne$^{22}$ abundances into better agreement.

WO stars are extremely C- and O-rich, as deduced from recombination analyses
(Kingsburgh, Barlow \& Storey 1995), and supported  by non-LTE models 
(Crowther et al. 2000). 
Further  nucleosynthesis reactions produce alpha elements via
\[
^{16}O (\alpha, \gamma) ^{20}Ne (\alpha, \gamma) ^{24}Mg (\alpha, \gamma) 
^{28} Si (\alpha, \gamma) ^{32} S 
 \]
producing a core which is initially dominated by $^{16}$O and $^{20}$Ne. 
{\it Spitzer} studies are in progress to determine neon abundances in WO 
stars, in order  to assess whether these stars show evidence of 
$\alpha$--capture  of oxygen (in which case enhanced $^{20}$Ne would again 
dominate  over $^{22}$Ne).

\section{Wind Properties}\label{mdot}

The existence of winds in early-type stars has been established since the
1960's, when the first rocket UV observations (e.g. Morton 1967) revealed the
characteristic P~Cygni signatures of mass-loss. Electron (Thompson) scattering
dominates the continuum opacity in O and WR stars, whilst the basic mechanism
by which their winds are driven is the transfer of photon momentum to the
stellar atmosphere through the absorption by spectral lines. The combination
of a plethora of spectral lines within the same spectral region as the
photospheric radiation allows for efficient driving of winds by radiation
pressure (Milne 1926).  Wind velocities can be directly measured, whilst 
wind density estimates  rely on varying complexity of theoretical 
interpretation. A
theoretical framework for mass-loss in normal hot, luminous stars was
developed by Castor, Abbott \& Klein (1975), known as CAK theory, via
line-driven radiation pressure.

\subsection{Wind velocities}

The wavelength of the blue edge of saturated P~Cygni absorption profiles
provides a measure of the asymptotic wind velocity. From these wavelengths,
accurate wind
velocities of WR stars can readily be obtained (Prinja, Barlow \& Howarth
1990; Willis et al. 2004).  Alternatively, optical and near-IR He\,{\sc i} 
P~Cygni profiles or mid-IR fine-structure metal lines   may be used to 
derive reliable wind velocities  (Howarth \& Schmutz 1992;  Eenens \& 
Williams 1994; Dessart et al. 2000).

In principle, optical recombination lines of He\,{\sc ii} and C\,{\sc 
iii-iv} may also be used to estimate  wind velocities, since these are 
formed close to the asymptotic flow velocity. However, velocities obtained 
from  spectral line 
modelling are preferable. For WR stars exhibiting weak winds -- whose lines 
are formed interior to the 
asymptotic flow velocity --  only lower velocity limits may be obtained. 
Nevertheless, observational evidence suggests lower wind
velocities at later subtypes by up to a factor of ten than early subtypes
(Table~\ref{WRtemp}).
Wind velocities of LMC WN stars compare closely with Milky Way counterparts.
Unfortunately, UV spectroscopy of SMC WN stars is scarce, such
that one has to rely on optical emission lines for wind velocity estimates,
which provide only lower limits to terminal wind velocities.

The current record holder amongst non-degenerate stars for the fastest stellar
wind is the Galactic WO star WR93b. For it, a wind velocity of 6000
km\,s$^{-1}$ has been obtained from optical recombination lines (Drew et al.
2004). Individual WO stars have now also been identified in a number of
external galaxies. One observes a reduction in line width (and so
wind velocity) for stars of progressively lower metallicity, by a factor of
up to two between the Milky Way and IC~1613 
(Crowther \& Hadfield 2006).  Although
numbers are small, this downward trend in wind velocity with decreasing
metallicity is believed to occur for other O and
WR  spectral types (e.g. Kudritzki \& Puls 2000; Crowther 2000).

\subsection{Mass-loss rates}

The mass-loss rate relates to the velocity field $v(r)$ and density
$\rho(r)$ via the equation of continuity
\begin{equation}\label{continuity} \dot{M} = 4 \pi r^2 \rho(r) v(r),
\end{equation} for a spherical, stationary wind.
WR winds may be observed at IR-mm-radio wavelengths via
the free-free (Bremsstrahlung) continuum excess caused by the 
stellar wind or via UV, optical or near-IR emission lines.

Mass-loss rates (e.g. Leitherer, Chapman \& Koribalski
1997) follow from radio continuum observations using
relatively simple analytical relations, under the
assumption of homogeneity and spherical symmetry.
The emergent radio flux $S_{\nu}$ depends on the
distance to the star $d$, mass-loss rate and terminal velocity as follows
\begin{equation} S_{\nu} \propto \left(
\frac{\dot{M}}{v_{\infty}}\right)^{4/3} \frac{\nu^{\alpha}}{d^2}
\end{equation} (Wright \& Barlow 1975), where $\alpha \sim 0.6$ 
in the constant-velocity regime. Accurate determinations of WR mass-loss 
rates depend upon composition, ionization balance and electron temperature 
at physical radii of $\sim 100-1000 R_{\ast}$.  Wind collisions  in an 
interacting binary system will cause additional non-thermal (synchrotron) 
radio  emission (Sect.~\ref{synch}), so care
needs to be taken against overestimating mass-loss rates in this way.

Optical spectral lines observed in WR stars can be considered as 
recombination lines, although line formation  is rather more complex 
in reality (Hillier 1988, 1989).  Since recombination involves the 
combination of ion and electron density, the strength of wind lines scales 
with the square of the density.  This explains why only a factor of 
$\sim$10 increase in wind
density with respect to the photospheric absorption line spectrum from O
supergiants produces an emission line Wolf-Rayet spectrum.  Recall the
comparison of $\zeta$ Pup (O4\,I(n)f) to HD~96548 (WR40, WN8) in
Figure~\ref{WR_ross}.

\subsection{Clumping}\label{clump}

There is overwhelming evidence in favour of 
highly clumped winds for WR and O stars. Line profiles
show propagating small-scale structures or `blobs', which are turbulent in 
nature (e.g. Moffat et al. 1988; L\'{e}pine et al. 2000). For optically
thin lines, these wind structures
have been investigated using radiation hydrodynamical simulations 
by Dessart \& Owocki (2005).  

Alternatively, individual  spectral 
lines, formed at  $\sim  10 R_{\ast}$, can be used to estimate
volume filling factors $f$ in WR winds 
(Hillier 1991). This technique permits an estimate of wind clumping factors
from a comparison between line electron 
scattering wings (which scale linearly with density)  and recombination 
lines (density-squared). This technique suffers from an 
approximate radial density 
dependence and is imprecise due to severe line blending, especially in WC stars. 
Nevertheless, fits to UV, optical and IR line profiles suggest
$f \sim 0.05-0.25$. As a consequence, 
global WR mass-loss rates are reduced by a
factor of $\sim 2-4$ relative to homogeneous models (dM/dt $\propto
f^{-1/2}$). Representative values are included in
Table~\ref{WRtemp}.  Spectroscopically derived mass-loss rates of
Milky Way WN stars span
a wide range of 10$^{-5.6}$ to 10$^{-4.4} M_{\odot}$ yr$^{-1}$. 
In contrast, Galactic WC stars
cover a much narrower range in mass-loss rate, 
from 10$^{-5.0}$ to 10$^{-4.4} M_{\odot}$ yr$^{-1}$.  

Independent methods support clumping-corrected WR mass-loss rates.
Binary systems permit use of the variation of linear polarization 
with orbital phase.  The modulation of linear polarization originates
from Thomson scattering of free electrons due to 
the relative motion of the companion with respect to the WR star.
This technique has been applied to several WR binaries
including V444 Cyg (HD~193576=WR139, WN5+O)  by St-Louis et al. (1993)
and has been developed further by Kurosawa, Hillier \& Pittard (2002)
using a Monte Carlo approach. For the case of V444~Cyg, polarization
results suggest a clumping factor of $f \sim 0.06$.

The most likely physical explanation for the structure in WR and O
star winds arises from theoretical evidence supporting 
an  instability in radiatively-driven winds (Lucy \& Solomon
1971; Owocki, Castor \& Rybicki 1988). There is a strong potential in
line scattering to drive wind material with accelerations that greatly
exceed the mean outward acceleration.  Simulations demonstrate that this
instability may lead naturally to structure. Such a flow is dominated by
multiple shock compressions, producing relatively soft 
X-rays. Hard X-ray fluxes from early-type stars are believed to 
be restricted to colliding wind binary
systems (e.g. $\gamma$ Vel: Schild et al. 2004), for which $10^{-7} \leq
L_{X}/L_{\rm bol} \leq 10^{-6}$.

\subsection{Metallicity dependent winds?}\label{metallicity}

We shall now consider empirical evidence in favour of
metallicity-dependent WR winds.
Nugis, Crowther \& Willis (1998) estimated mass-loss
rates for Galactic WR stars from archival radio observations, allowing
for clumped winds. Nugis \& Lamers (2000) provided 
empirical mass-loss scaling relations by
adopting physical parameters derived from spectroscopic analysis and/or
evolutionary predictions. For a combined sample of WN and WC stars,
Nugis \& Lamers (2000)
obtained
\begin{equation}\label{nugis-lamers00} \log \dot{M}/(M_{\odot} {\rm
yr}^{-1}) = -11.00 + 1.29 \log L/L_{\odot} + 1.74 \log Y + 0.47 \log Z
\end{equation} where Y and Z are the mass fractions of helium and metals,
respectively.

\subsubsection{WN winds}

\begin{figure}
\centerline{\psfig{file=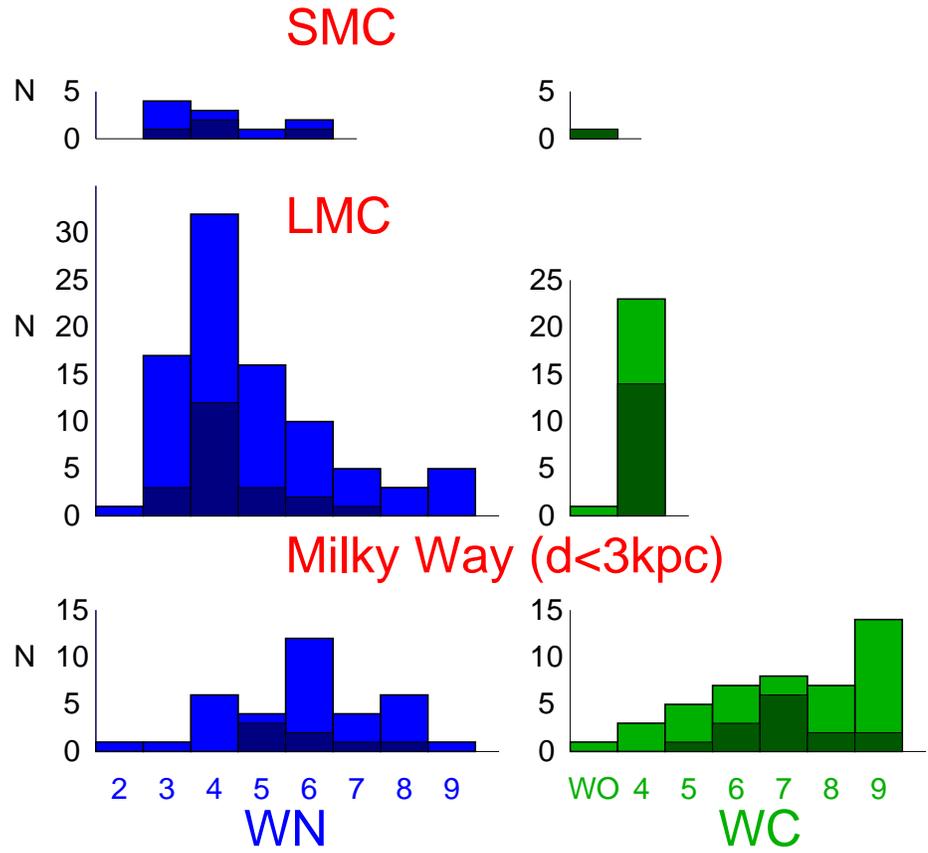,width=4.8in}}
\caption{Subtype distribution of Milky Way ($d<$3kpc), 
LMC, and SMC WR stars, according to van der Hucht (2001), Bartzakos, Moffat 
\& Niemela (2001), Foellmi, Moffat  \& Guerrero (2003ab). Both 
visual and close WR  binaries are shaded (e.g. only 3 of the LMC WC4 stars 
are close binaries according to Bartzakos et al. 2001). Rare, intermediate 
WN/C stars  are included in the WN sample.}
\label{wrpop}
\end{figure}

\begin{figure}
\centerline{\psfig{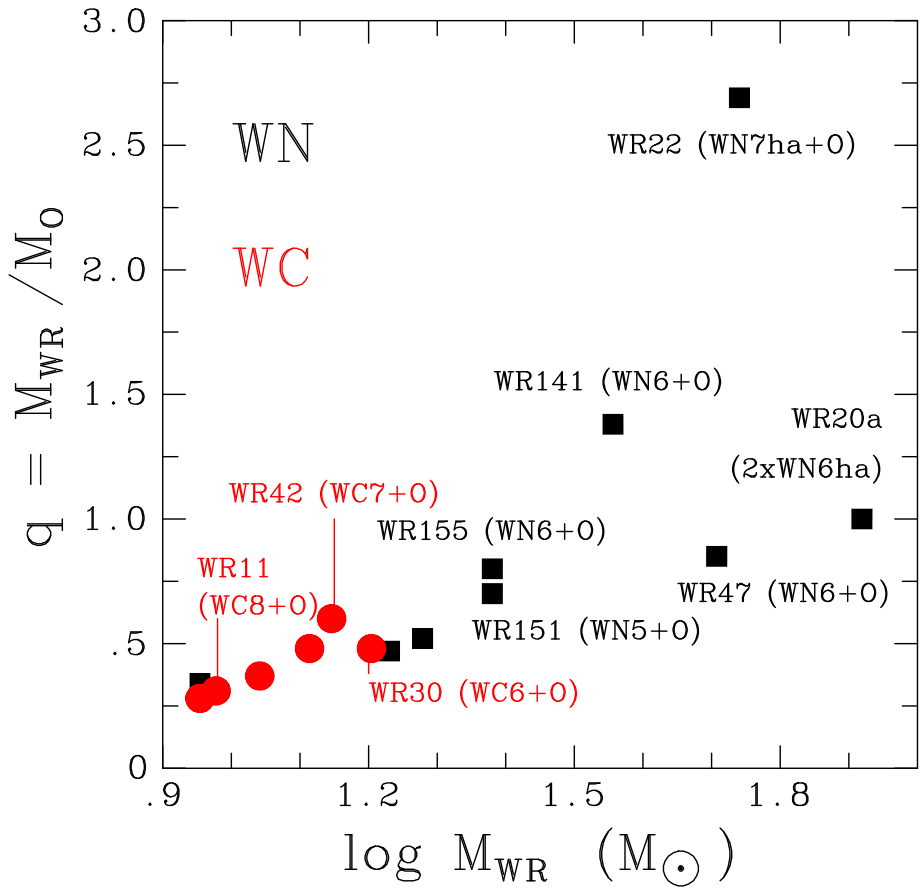}}
\caption{Stellar masses for Milky Way WR stars obtained from binary orbits 
(van der Hucht 2001; Rauw et al. 2005; Villar-Sbaffi 
et al. 2006)}
\label{binary_masses}
\end{figure}

\begin{figure}
\centerline{\psfig{file=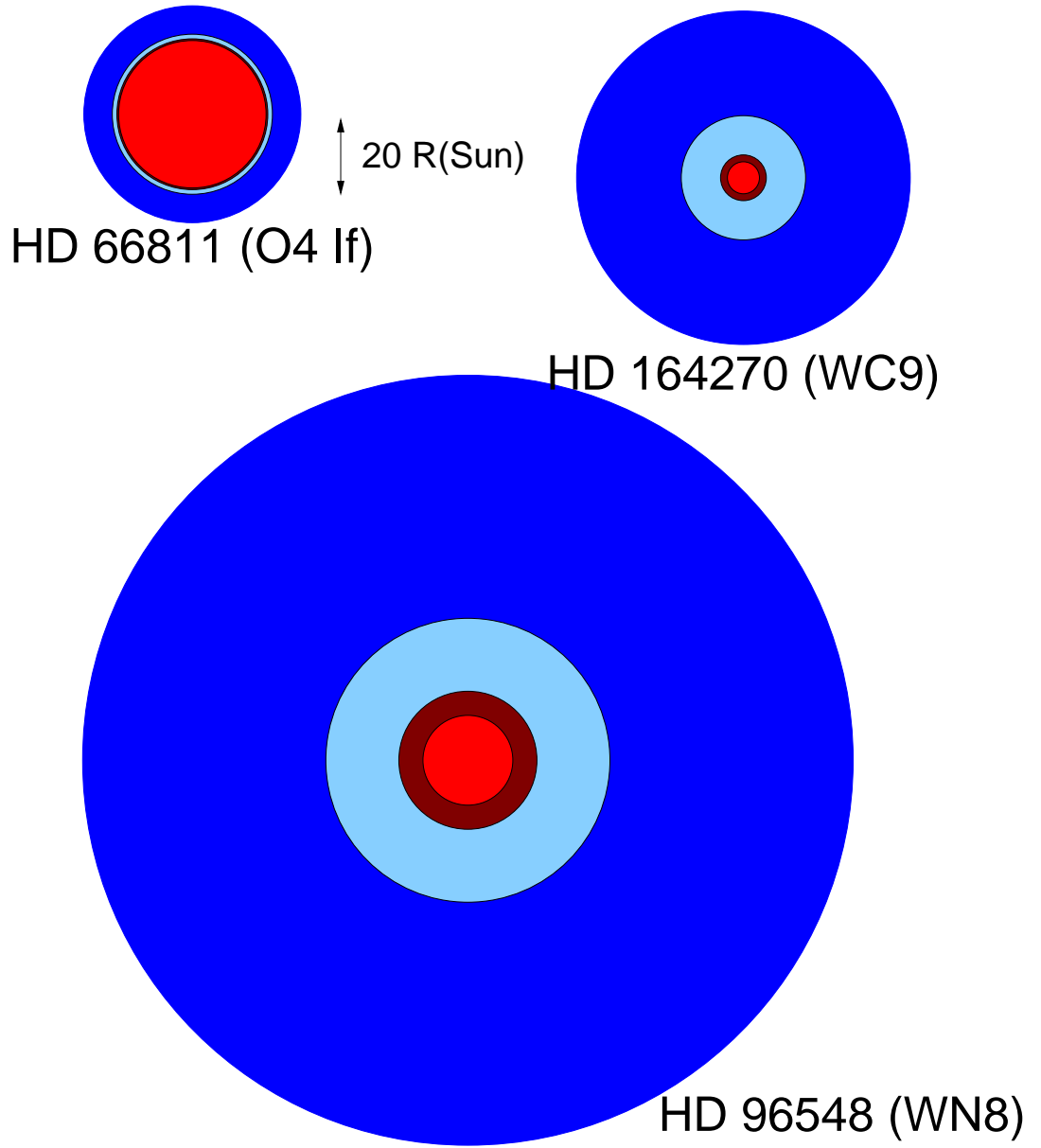,width=15cm}}
\caption{Comparisons between stellar radii at Rosseland 
optical depths of 20 (= $R_{\ast}$, red) and 2/3 (= $R_{2/3}$, dark
red) for HD~66811 (O4\,If), HD~96548 (WR40, WN8) and HD~164270 
(WR103, WC9), shown 
to scale. The primary
optical wind line-forming region, $10^{11} 
\leq n_{e} \leq 10^{12}$ 
cm$^{-3}$ is shown in dark blue, plus higher density wind material,
$n_{e} \geq 10^{12}$ cm$^{-3}$ is indicated in light
blue. The figure illustrates the highly extended
winds of WR stars with respect to O stars (Repolust, Puls \& Herrero
2004; Herald, Hillier
\& Schulte-Ladbeck 2001; Crowther, Morris \& Smith 2006a).}
\label{WR_ross}
\end{figure}

\begin{figure}
\centerline{\psfig{file=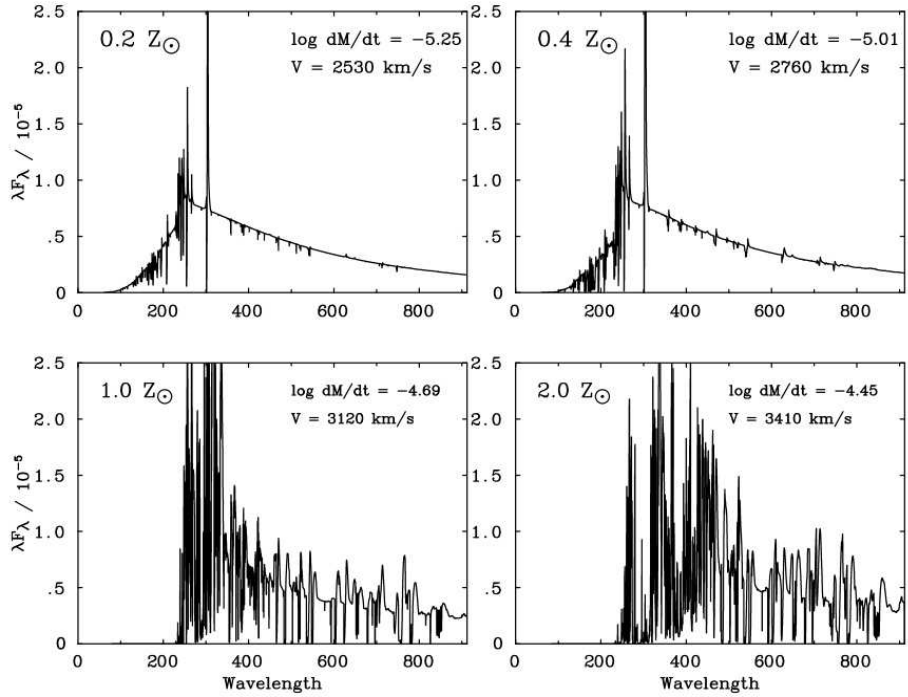,width=4.8in,angle=0}}
\caption{Comparison between the Lyman continuum 
ionizing fluxes of early WN CMFGEN models with fixed parameters
(100\,kK, $\log L/L_{\odot}$ = 5.48), except that the mass-loss rates 
and wind velocities depend upon metallicities according to Smith, 
Norris \& Crowther (2002). Only the low wind density models 
predict a significant flux below the He$^{+}$ edge at 228\AA}
\label{wn_grid}
\end{figure}

Smith \& Willis (1983) compared the properties of WN stars in the LMC
and Milky Way, concluding there was no significant differences between the 
wind properties of the two samples. 
These conclusions were supported by Hamann \& Koesterke (2000) from
detailed non-LTE modelling, although a large scatter in mass-loss rates
within each parent galaxy was revealed. Either there is
no metallicity dependence, or any differences are too subtle to be
identified from the narrow metallicity range spanned by the Milky Way and
LMC. Within 
the Milky Way, most late-type stars contain hydrogen and most early-type
stars do not. In contrast, 
early-type stars dominate WN populations in the Magellanic Clouds
(recall Figure~\ref{wrpop}), and
often contain atmospheric hydrogen
(Smith, Shara \& Moffat 1996; Foellmi, Moffat \& 
Guerrero 2003ab).

\begin{figure}
\centerline{\psfig{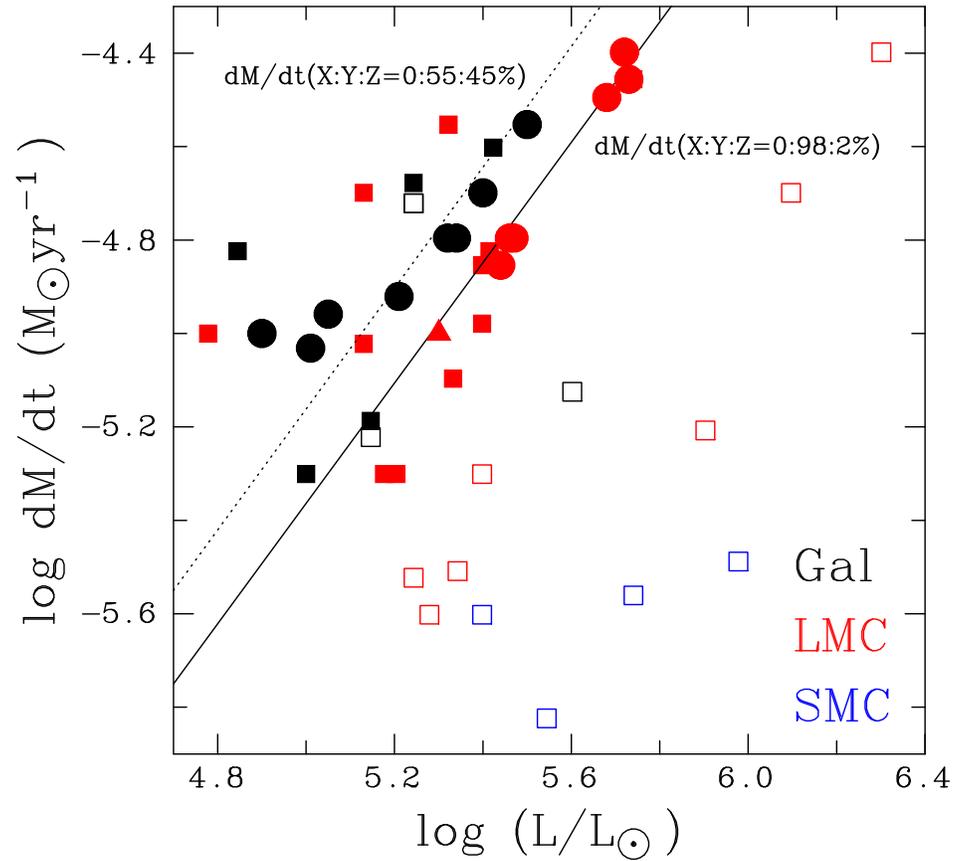}}
\caption{Comparison between the mass-loss rates and
luminosities of WN3--6 (squares), WC5--9 (circles) and WO (triangles) stars 
in the Galaxy (black), LMC (red) and SMC (blue). 
Eqn~\ref{nugis-lamers00} 
from  Nugis \& Lamers (2000) for H-poor WN (solid line)  and WC
stars (dotted line) is included. Open/filled symbols refer to 
WN stars with/without surface hydrogen, based upon analysis 
of  near-IR  helium lines (Crowther 2007). 
Mass-loss rates are universally high if hydrogen is absent.}
\label{mdot_plot}
\end{figure}

Figure~\ref{mdot_plot} compares the mass-loss rates
of cluster or association member WN stars in the Milky Way with
Magellanic Cloud counterparts. Mass-loss estimates are 
obtained from their near-IR helium lines 
(Crowther 2007, following Howarth \& Schmutz 1992). 
The substantial scatter in mass-loss rates is in line with the 
heterogeneity of line strengths within WN subtypes. 
Stronger winds are measured for WN 
stars without surface hydrogen, in agreement
with Eqn~\ref{nugis-lamers00}) and 
recent results of Hamann, Gr\"{a}fener \& Liermann (2006). 
From the Figure, measured mass-loss rates of 
hydrogen-rich early-type WN stars in the SMC (1/5~$Z_{\odot}$) are 
0.4 dex weaker than equivalent stars in the Milky Way and LMC 
(1/2 to 1~$Z_{\odot}$). This suggests a metallicity dependence of 
dM/dt $\propto Z^{m}$ for WN stars, with $m \sim 0.8\pm0.2$. 
The exponent is comparable to that measured from H$\alpha$
observations of Milky Way, LMC and SMC O-type stars (Mokiem et al. 2007). 
 
There are two atmospheric factors which contribute to the observed
trend towards earlier WN subtypes at lower  metallicities; 
\begin{enumerate}
\item CNO compromises $\sim$1.1\% by mass of the Solar photosphere (Asplund et 
al. 2004) versus 0.48\% in the LMC and 0.24\% in the SMC (Russell \& Dopita
1990). Since WN stars typically exhibit CNO equilibrium abundances, there 
is a maximum nitrogen content available in a given environment. For otherwise
identical physical parameters, Crowther (2000) demonstrated that a reduced
nitrogen content at lower metallicity favours an earlier subtype. This is
regardless of metallicity dependent mass-loss rates, and results from 
the abundance sensitivity of nitrogen classification lines.
\item Additionally, a metallicity dependence of WN winds would
enhance the trend to earlier spectral subtypes. Dense WN winds
at high metallicities lead to 
efficient recombination from high ionization stages (e.g. 
N$^{5+}$) to lower ions (e.g. N$^{3+}$) within the optical line 
formation regions.  This would not occur so effectively for low density 
winds, enhancing the trend towards early-type WN stars in metal-poor
environments.
\end{enumerate} Consequently, both effects favour predominantly late 
subtypes at  high metallicity, and early  subtypes at low metallicity, 
which is indeed generally observed (Fig.~\ref{wrpop}).

\subsubsection{WC winds}

It is well established that WC stars in the inner Milky Way, and indeed
all metal-rich environments, possess later spectral types than those in
the outer Galaxy, LMC and other metal-poor environments
(Figure~\ref{wrpop}; Hadfield et al. 2007). 
This observational trend led Smith \& Maeder (1991) to suggest
that early-type WC stars are richer in carbon than late-type WC stars,
on the basis of tentative results from recombination line analyses.
In this scenario, typical Milky Way  WC5--9 stars exhibit reduced
carbon abundances than WC4 counterparts in the LMC. However, 
quantitative analysis of WC subtypes allowing for radiative transfer effects 
do not support a subtype dependence of elemental abundances in WC stars
 (Koesterke \& Hamann 1995), as discussed in Sect~\ref{abundances}.

If differences of carbon content are not responsible for the observed WC
subtype distribution in galaxies, what is its origin?
Let us consider the WC classification lines -- C\,{\sc iii} $\lambda$5696 and
C\,{\sc iv} $\lambda\lambda$5801-12 -- in greater detail. Specifically the upper level of
$\lambda$5696 has an alternative decay via $\lambda$574, with a branching
ratio of 147:1 (Hillier 1989). Consequently $\lambda$5696 only becomes 
strong when
$\lambda$574 is optically thick, i.e. if the stellar temperature is low
{\it or} the wind density is sufficiently high. 
From non-LTE models it has been established that the temperatures of Galactic 
WC5--7 stars and LMC WC4 stars are similar, such 
that the observed subtype distribution argues that the wind densities of 
Galactic WC stars must be higher than the LMC stars.

Figure~\ref{mdot_plot} also compares clumping corrected 
mass-loss rates of  WC stars in the Milky Way and LMC, as derived
from optical studies. The Galactic sample agree 
well with Eqn~\ref{nugis-lamers00} from Nugis \& Lamers (2000).
Crowther et al. (2002) obtain a similar mass-loss dependence for WC4
stars in the LMC, albeit offset by --0.25 dex. The comparison between
derived LMC and Solar neighbourhood WC wind properties suggests a 
dependence of dM/dt $\propto Z^{m}$ with $m \sim 0.6 \pm 0.2$. 
Crowther et al. (2002) argued that the WC subtype distributions 
in the LMC and Milky Way resulted from this metallicity dependence.
C\,{\sc iii} $\lambda$5696 emission is very sensitive to mass-loss rate, so 
weak winds for LMC WC stars would produce negligible C\,{\sc iii}
$\lambda$5696 emission (WC4 subtypes) and strong winds interior to
the Solar circle would produce strong C\,{\sc iii} $\lambda$5696 emission
(WC8-9 subtypes), in agreement with the observed subtype distributions.

\subsection{Line driving in WR winds}\label{line_driving}

Historically, it has not been clear whether radiation pressure alone is
sufficient to drive the high mass-loss rates of WR stars. Let us
briefly review the standard Castor, Abbott \& Klein (1975, hereafter CAK)
theory behind radiatively driven winds before addressing the question of
line driving for WR stars. Pulsations have also been proposed for WR 
stars, as witnessed in intensive monitoring for  the most photometrically 
variable WN8 star with the {\it MOST} satellite by Lef\'{e}vre et al. 
(2005). Interpretation of 
such observations however remains ambiguous (Townsend \& MacDonald 2006; 
Dorfi, Gautschy \& Saio 2006).

\subsubsection{Single scattering limit}

The combination of plentiful line opacity in the extreme UV, where the 
photospheric radiation originates, allows for efficient driving of hot 
star winds by radiation pressure (Milne 1926). In a static atmosphere, the 
photospheric radiation will only be efficiently absorbed or scattered in 
the lower layers of the atmosphere, weakening the radiative acceleration, 
$g_{\rm line}$, in the outer layers. In contrast, atoms within the outer 
layers of an expanding atmosphere see the photosphere as Doppler-shifted 
radiation, allowing absorption of undiminished continuum photons in their 
line transitions (Sobolev 1960).

The force from optically thick lines, which provide the radiative 
acceleration by absorbing the photon momentum, scale with the velocity 
gradient. There can be at most $\approx c/v_{\infty}$ thick 
lines, implying a so-called single-scattering limit
\begin{equation}
\dot{M} v_{\infty}  \leq L/c.
\end{equation}
Castor, Abbott \& Klein (1975) and Abbott (1982)
developed a self-consistent solution of the wind 
properties, from which one obtains a velocity `law'
\begin{equation}\label{velocity_law}
v(r) = v_{\infty} \left( 1 - \frac{R_{\ast}}{r} \right)^{\beta},
\end{equation}
for which $\beta$=0.8 for O-type stars (Pauldrach, Puls \& Kudritzki 
1986). $g_{\rm line}$ can be written in terms of the
Thompson (electron) scattering acceleration, i.e. $g_{\rm line} = g_{e} 
X(t)$. The equation of motion can then be expressed as
\begin{equation}
v \frac{dv}{dr} = \frac{GM}{r^{2}} \left(\Gamma_{e} (1+ X(t)) -1\right).
\end{equation}
It is clear that both $\Gamma_{e}$ and $X(t)$ need to 
be large for a hot star to possess a wind.
In the CAK approach, optically thick lines are assumed not to overlap
within the wind. In reality, this is rarely true in the extreme UV where
spectral lines are very tightly packed and the bulk of the line driving
originates. Consequently, another approach is 
needed for WR stars whose winds exceed the single scattering  limit 
(Lamers \&  Leitherer 1993), namely the consideration of
multiple scattering.

\subsubsection{Multiple scattering}

Historically, the strength of WR  winds 
were considered to be mass dependent, but metallicity-independent
(Langer 1989). Observational evidence now favours  metallicity-dependent 
WR winds, with a dependence of dM/dt $\propto Z^{m}$, with $m \sim 0.8$ 
for WN stars (Sect.~\ref{metallicity}). It is established that
O star winds are driven by radiation pressure, with a metallicity 
dependence that is similar to WN stars (Mokiem et al. 2007). Consequently,
the notion that WR winds are  radiatively driven is 
observationally supported.

Theoretically, Lucy \& Abbott (1993) and Springmann (1994) produced Monte 
Carlo wind models for WR stars in which multiple-scattering was achieved 
by the presence of multiple  ionization stages in the wind. 
However, a prescribed velocity 
and ionization  structure was  adopted in both  case studies, 
plus the inner wind acceleration was not explained. 

Schmutz (1997) first tackled the problem of driving WR winds from
radiatively-driven winds self-consistently using a
combined Monte Carlo and radiative transfer approach. 
He also introduced a
means of photon-loss from the He\,{\sc ii} Ly$\alpha$ 303\AA\ line via a 
Bowen resonance-fluorescence mechanism. This effect led to a change in 
the ionization equilibrium of helium, requiring a higher stellar 
luminosity. The consideration of wind clumping by Schmutz (1997) 
did succeed in providing a sufficiently strong outflow  in the outer wind. 
Photon-loss nevertheless failed to initiate the requisite
powerful acceleration in the deep atmospheric layers. Subsequent studies  
have  supported the principal  behind the  photon-loss mechanism for WR  
stars, although the effect is modest (e.g. De Marco et al. 2000). 

The next advance for the inner wind driving was by Nugis \& Lamers (2002) 
whose analytical study suggested that the (hot) iron opacity  peak 
at 10$^{5.2}$~K is responsible for the observed WR mass-loss in an 
optically thick wind (a cooler opacity peak exists at 10$^{4.6}$K).
Indeed, Gr\"{a}fener \& Hamann  (2005) established that highly ionized Fe 
ions (Fe\,{\sc  ix-xvii}) provides the necessary opacity for initiating
WR winds deep in the atmosphere for WR111  
(HD~165763=WR111, WC5). The wind acceleration due to radiation and  gas  
pressure self-consistently matches the mechanical and gravitational 
acceleration in their hydrodynamical model. Gr\"{a}fener \& Hamann  (2005) 
achieved the observed terminal wind velocity by adopting an extremely low 
outer wind filling factor of $f$=0.02. This degree of clumping is
unrealistic since predicted line  electron scattering wings are too 
weak with respect to  observations. A more physical outer wind solution
should be permitted by the inclusion of  more complete 
opacities from other 
elements such as Ne and Ar. The velocity structure from the Gr\"{a}fener 
\& Hamann (1995) hydrodynamical model 
closely matches a  typical $\beta$=1 velocity law of the form in 
Eqn~\ref{velocity_law} in the inner wind. A slower $\beta$=5 law is
more appropriate for the outer  wind. Indeed, such a hybrid velocity 
structure was first proposed by Hillier \&  Miller (1999).

Theoretically, both the hydrodynamical models of Gr\"{a}fener \& Hamann
(2005) and recent Monte Carlo wind models for WR stars by Vink \& de
Koter (2005) argue in favour of radiation pressure through metal lines as
responsible for the observed multiple-scattering in WR winds. 
The critical parameter involving the  development of strong outflows
is the proximity of WR stars to the Eddington limit, according to 
Gr\"{a}fener \& Hamann (2007). 

Vink \& de Koter (2005) performed a multiple-scattering study of WR stars
at fixed stellar temperatures and Eddington parameter. A metallicity 
scaling of dM/dt $\propto Z^{m}$ with $m$=0.86 for 10$^{-3} \leq 
Z/Z_{\odot} \leq 1$ was obtained. This exponent is similar to empirical WN 
and O star results across a more restricted metallicity range.
Gr\"{a}fener \& Hamann (2007) predict a decrease in the exponent
or late-type WN stars at higher $\Gamma_{e}$, plus reduced wind velocities
at lower metallicities. Vink \& de Koter (2005) predict that the 
high metal content of WC stars favour a weaker dependence with metallicity 
than WN stars. A mass-loss scaling with exponent
$m$=0.66 for  $10^{-1} \leq Z/Z_{\odot} \leq 1$ is predicted. This
is consistent with the observed WC mass-loss dependence between the LMC
and Milky Way. At low metallicity, Vink \& de Koter (2005) predict
a weak dependence of $m$=0.35 for $10^{-3} \leq 
Z/Z_{\odot} \leq 10^{-1}$ providing atmospheric carbon and oxygen abundances
are metallicity-independent.

\section{Interacting Binaries}

\subsection{Close binary evolution}

The components of a massive binary may evolve independently, as if they
were single stars providing their orbital periods are sufficiently large. 
In contrast, an interacting or close binary system represents the case in
which the more massive primary component 
expands to fill its Roche lobe, causing
mass exchange to the secondary. For an initial period of several days,
the primary will reach its Roche lobe whilst still on the H-core burning
main sequence (Case A) and transfer the majority of its hydrogen-rich
envelope to the secondary. For initial periods of a few weeks or years,
mass transfer will occur during the hydrogen-shell burning phase (Case B)
or He-shell burning (Case C), respectively. Cases B and C are much more common
than Case A due to the much larger range of orbital periods sampled. 
According to Vanbeveren, De Loore \& Van Rensbergen
(1998) a common envelope will occur instead of Roche lobe overflow if
there is an LBV phase.

Close binary evolution will extend the formation of WR stars to lower initial mass, 
and consequently lower luminosity with respect to single star evolution 
(e.g. Vanbeveren et al. 1998). The
observed lower mass limit to WR formation in the Milky Way is broadly
consistent with the single star scenario (Meynet \& Maeder 2003). Therefore,
either low-mass He stars are not recognised spectroscopically as WR stars, 
or they may be too heavily diluted by their brighter O star companions 
for them to be observed.

One natural, albeit rare, consequence of massive close binary evolution
involves the evolution of the initial primary to core-collapse, with a
neutron star or black hole remnant, in which the system remains bound as
a high mass X-ray binary (HMXB, Wellstein \& Langer 1999). The OB
secondary may then evolve through to the Wolf-Rayet phase, producing a WR
plus neutron star or black hole binary. For many years, searches for such
systems proved elusive, until it was discovered that Cyg X--3, a 4.8 hour period
X-ray bright system possessed the near-IR spectrum of a He-star (van Kerkwijk et al.
1992). Nevertheless, the nature of Cyg X--3 remains somewhat
controversial. WR plus compact companion candidates have also been 
identified in external galaxies, IC~10 X--1 (Bauer \& Brandt 2004) 
and NGC~300 X-1 (Carpano et al. 2007).

\subsection{Colliding winds}\label{synch}

The presence of two early-type stars within a binary
system naturally leads to a wind-wind interaction region. In general, details
of the interaction process are investigated by complex hydrodynamics.
Nevertheless, the analytical approach of Stevens, Blondin \& Pollock
(1992) provides a useful insight into the physics of the colliding winds.

A subset of WR stars display  non-thermal (synchrotron) radio
emission, in addition to the thermal radio emission produced via free-free
emission from their stellar wind. Consequently, a magnetic field must
be present in the winds of such stars, with relativistic electrons
in the radio emitting region. Shocks associated with a wind
collision may act as sites for particle acceleration through the 
Fermi mechanism (Eichler \& Usov 1993). Free electrons would
undergo acceleration to relativistic velocities by crossing the shock
front between the interacting stellar winds. Indeed,
the majority of non-thermal WR radio emitters are known
binaries. For example, WR140 (WC7+O) is a
highly eccentric system with a 7.9 year period, in which the
radio flux is thermal over the majority of the orbit. Between
phases 0.55 and 0.95 (where phase 0 corresponds to periastron passage), 
the radio flux increases dramatically and displays a non-thermal radio 
index (Williams et al. 1990).

One may introduce the wind momentum ratio, $R$
\begin{equation} 
R = \frac{\dot{M}_{1} v_{\infty,1}}{\dot{M}_{2}
v_{\infty,2}} 
\end{equation} where the mass loss rates and wind velocities
of components $i$ are given by $\dot{M_{i}}$ and $v_{\infty,i}$,
respectively. In the simplest case of $R$=1, the intersection between the
winds occurs in a plane midway between the two stars. In the more likely
situation of $R \neq$ 1, the contact discontinuity appears as a cone
wrapped around the star with the less momentum in its  wind. 
By way of example, radio emission in the WR147 (WN8 + B0.5) system has been 
spatially resolved into two
components: thermal emission from the WN8 primary wind, plus non-thermal
emission located close to the companion (Williams et al. 1997). The
location of the non-thermal component is consistent with the ram pressure
balance of the two stellar winds,
from which a momentum ratio of $R \sim 0.011$ can be obtained. 

For early-type stars that have reached their terminal velocities of
several thousand km\,s$^{-1}$, the post-shock plasma temperature is very 
high
($\geq 10^{7}$~K). X-rays represent the 
main observational signature of shock-heated plasma. Significantly harder
X-ray fluxes are expected from massive binaries with respect to 
single stars. Phase-locked X-ray variability is expected due to the
change in opacity along the line-of-sight, or varying separation
for eccentric binaries (Stevens, Blondin \& Pollock 
1992). A colliding wind system in which substantial
phase-locked variability has been observed with {\it ROSAT} is 
$\gamma$ Vel (WC8+O). The X-ray emission from the shock is absorbed
when the opaque wind from the WR star lies in front of the O star. When the
cavity around the O star crosses our line-of-sight, X-ray emission
is significantly less absorbed (Willis, Schild \& Stevens 1995).

\subsection{Dust formation}

The principal sources of interstellar dust are cool, high mass-losing
stars, such as red giants, asymptotic giant branch stars, plus novae and
supernovae. Dust is observed around some massive stars, particularly LBVs
with ejecta nebulae, but aside from their giant eruptions, this may be
material that has been swept up by the stellar wind. The intense radiation
fields of young, massive stars would be expected to prevent dust formation 
in their local environment. However, Allen, Swings \& Harvey (1972) identified
excess IR emission in a subset of WC stars, arising from 
$\sim$1000~K circumstellar dust.

Williams, van der Hucht \& Th\'{e} (1987)  investigated the infrared
properties of Galactic WC stars, revealing persistent dust formation in
some systems, or episodic formation in other cases. For a single star
whose wind is homogeneous and spherically symmetric, 
carbon is predicted to remain singly or even doubly ionized due to high
electron temperatures of $\sim$10$^{4}$K in the region where dust
formation is observed to occur. However, the formation of graphite or 
more likely amorphous carbon grains requires a high density of neutral 
carbon close to the WC star.

One clue to the origin of dust is provided
by WR140 (HD~193793, WC7+O) which forms dust episodically, near periastron
passage. At this phase the power in the colliding winds is at its
greatest (Williams et al. 1990).  Usov (1991) analytically showed
that the wind conditions of WR140 at periastron favour a strong gas
compression in the vicinity of the shock surface, providing an
outflow of cold gas. It is
plausible that high density, low temperature, carbon-rich material 
associated with the bow-shock in a colliding wind WC binary 
provides the necessary environment for dust formation.

In contrast to episodic dust formers, persistent WC systems are rarely
spectroscopic WC binaries, for which WR104 (Ve~2-45, WC9) is the prototype 
identified by Allen et al. (1972). Spectroscopic evidence from Crowther
(1997) suggested the presence of an OB companion in the WR104 system
when the inner WC wind was obscured by a dust cloud, analogous to R~Coronae
Borealis stars. 
Conclusive proof of the binary nature of WR104 has been
established by Tuthill, Monnier \& Danchi (1999)  from high spatial
resolution near-IR imaging.  Dust associated with WR104 forms a spatially
confined stream that follows a spiral trajectory (so-called `pinwheel'),
analogous to a garden rotary sprinker. The cocoon stars after which the
Quintuplet cluster at the Galactic Centre was named have also been
identified as dusty WC pinwheel stars by Tuthill et al.  (2006).

Binarity appears to play a key role in the formation of dust in WC stars,
providing the necessary high density within the shocked wind interaction
region, plus shielding from the hard ionizing photons. 
The presence of hydrogen from the OB companion may provide
the necessary chemical seeding in the otherwise hydrogen-free 
WC environment. Alas, this possibility does have difficulties, since chemical
mixing between the WC and OB winds may not occur in the immediate
vicinity of the shock region. Nevertheless, it is likely that
all dust forming WC stars are binaries.

\section{Evolutionary models and properties at core-collapse}

The various inputs to stellar interior evolutionary models originate from
either laboratory experiments (e.g. opacities, nuclear reaction rates) or
astronomical observations (e.g. mass-loss properties, rotation rates).
Indeed, mass-loss (rather than convection) has a dominant effect upon 
stellar models for the most massive stars. 

Here we illustrate one of the potential pitfalls of this approach.
According to Koesterke et al. (1991), stellar luminosities of
some weak-lined early-type WN stars were unexpectedly low ($\leq 10^{5} 
L_{\odot}$).
In order to reproduce such results, plus the observed N(WR)/N(O) ratio,
Meynet et al. (1994) adopted higher stellar mass-loss rates, with respect to
previous empirical calibrations. Improvements in  
non-LTE models  have subsequently led to higher derived 
stellar luminosities (e.g. Hamann, Gr\"{a}fener \& Liermann 2006). Revised
WR luminosities removed the primary motivation behind elevated 
mass-loss rates.  Indeed, allowance for wind clumping has now
led to the need to reduce mass-loss rates in evolutionary models.

\subsection{Rotational mixing}

Mass-loss and rotation are intimately linked for the evolution of massive 
stars. Stellar winds will lead to spin-down for the case of an efficient 
internal angular momentum transport mechanism. At Solar metallicity, one 
anticipates rapid spin-down for very massive stars due to their strong 
stellar winds (Langer 1998). Initial rotational velocities are erased 
within a few million years. In contrast, initial conditions may remain 
preserved throughout the main sequence lifetime of O-type stars in 
metal-poor environments due to their weak stellar winds.

Discrepancies between evolutionary model predictions and a number of 
observed properties of high mass stars led to the incorporation of 
rotational mixing into interior models, following the theoretical 
treatment of Zahn (1992). Rotational mixing reproduces some of the 
predictions from the high mass-loss evolutionary models of Meynet et al.  
(1994), for which two approaches have been developed. Meynet \& Maeder 
(1997)  describe the transport of angular momentum in the stellar interior 
through the shear and meridional instabilities. In contrast, momentum is 
transported radially from the core to the surface in the approach of Heger 
et al. (2000).

Rotation favours the evolution into the WR phase at earlier stages, 
increasing the WR lifetime. Lower initial mass stars also enter the
WR phase. For an assumed initial rotational velocity of 300~km\,s$^{-1}$,
the minimum initial mass star entering the WR phase is 22~$M_{\odot}$,
versus 37$M_{\odot}$ for non-rotating models at Solar metallicity 
(Meynet \& Maeder 2003). Evolutionary models allowing for rotational
mixing do predict a better agreement with the observed  ratio of WR to O 
stars at low metallicity, the existence of  intermediate WN/C stars 
(though see Langer  1991), and the ratio of blue to red supergiants in
galaxies.

Regarding the initial rotation velocities of massive stars, evolutionary 
models adopt fairly high values. Observationally, $v_{\rm rot} \sim 175 \pm 
125$ km\,s$^{-1}$ for young OB stars in the SMC cluster NGC~346 (Mokiem et 
al. 2006), suggesting somewhat lower initial rotation rates, on average.  
Nevertheless, lower mass limits to the formation of WR stars are predicted 
to lie in the range 42$M_{\odot}$ at Z=0.004 (SMC) to 21$M_{\odot}$ at 
Z=0.04 (M~83) for models allowing for such high initial rotation rates 
(Meynet \& Maeder 2005). Hirschi, Meynet \& Maeder (2005) present chemical 
yields from rotating stellar models at Solar metallicity, revealing 
increased C and O yields below 30$M_{\odot}$, and higher He yields at 
higher initial masses.

\subsection{Evolutionary model predictions}

It is possible to predict the number ratio of WR to O stars for
regions of constant star formation from rotating evolutionary 
models, weighted over the Initial Mass Function (IMF).  
For an assumed Salpeter IMF slope for
massive stars, the ratios predicted are indeed in
much better agreement with the observed distribution at
Solar metallicity (Meynet \& Maeder 2003). Since the O star
population is relatively imprecise, the predicted WR subtype distributions
are often used instead for comparisons with observations.

From Figure~\ref{wrpop}, the Solar Neighbourhood WR subtype distribution
contains similar numbers of WC and WN stars, with an equal number 
of early (H-free) and late (H-rich) WN stars. From comparison with
evolutionary models, the agreement is reasonable, except for the brevity
of the H-deficient WN phase in interior models at Solar metallicity. This
aspect has been quantified by Hamann, Gr\"{a}fener \& Liermann (2006). 
Synthetic WR populations from the Meynet \& Maeder (2003)
evolutionary tracks predict 
that only 20\% of WN stars should be hydrogen-free, in
contrast to over 50\% of the observed sample.  Non-rotating models
provide better statistics, although low luminosity early-type WN
stars are absent in such synthetic populations.

\begin{figure}
\centerline{\psfig{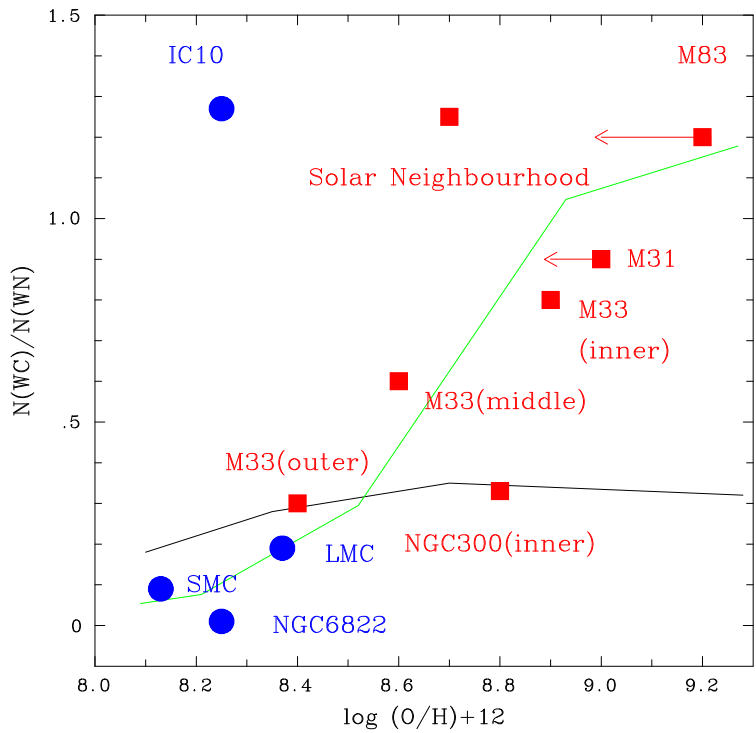}}
\caption{Comparison between observed N(WC)/N(WN) ratio and oxygen content,
for nearby spiral (red) and irregular (blue) galaxies (Massey \& Johnson
1998; Crowther et al. 2003; Schild et al. 2003; Hadfield et al. 2005)
together with evolutionary model  predictions by Meynet \& Maeder (2005, 
black) and Eldridge \& Vink (2006, green).  Different regions of M33
are shown (inner, middle, outer), resulting from the strong metallicity 
gradient in that galaxy.}
\label{wcwn}
\end{figure}

Figure~\ref{wcwn} shows that the ratio of WC to WN stars is observed to 
increase with metallicity for nearby galaxies whose WR content has been 
studied in detail. One notably exception is the low-metallicity Local 
Group galaxy IC~10 (Massey \& Holmes 2002; Crowther et al. 2003). The WR 
population of IC~10 remains controversial, since high Galactic foreground 
extinction favours the detection of WC stars over WN stars. The 
preferential detection of WC stars arises because the equivalent widths of 
the strongest optical lines in WC stars are (up to 100 times) larger than 
those of the strongest optical lines in WN stars (Massey \& Johnson 1998).

Two evolutionary model predictions are included in Fig.~\ref{wcwn}; (a)
allowing for rotational mixing but without a WR metallicity 
scaling (Meynet \& Maeder 2005); (b) neglecting rotational mixing, 
although with a 
metallicity scaling for WR stars (Eldridge \& Vink 2006). The latter 
models, in which convective overshooting is included, agree better 
with observations at higher metallicities. It should be emphasised that 
a significant WR population formed via a close binary channel 
is required to reproduce the observed WR/O ratio across
the full metallicity range in the Eldridge \& Vink (2006) models
(see also Van Bever \& Vanbeveren 2003). A significant binary channel
is not required for the Meynet \& Maeder (2005) rotating 
evolutionary models. These 
resolve many issues with  respect to earlier comparisons to observations, 
although some problems persist. 

In very metal-poor environments ($\sim$1/50 $Z_{\odot}$) the WR
phase is predicted for only the most massive single stars ($\geq90
M_{\odot}$)  according to non-rotating models of de Mello et al. (1998).
Nevertheless, WR stars have been observed in UV and optical spectroscopy 
of such metal-poor regions within I\,Zw~18 (Izotov et al. 1997; Brown et 
al. 2002) and
SBS~0335-052E (Papaderos et al. 2006).  Only WC stars have been
unambiguously identified spectroscopically, yet WN stars would be 
expected to dominate the WR population of such metal-poor regions. 
The strength of WN winds  
are believed to depend more sensitively upon metallicity than the
strength of WC winds
(Vink \& de Koter 2005). Therefore, WN stars may be extremely 
difficult to directly detect. Hot
weak-lined WN stars are predicted to have hard UV ionizing flux
distributions, so they may be indirectly indicated via the presence of
strong nebular He\,{\sc ii} $\lambda$4686 emission. Indeed, strong nebular
He\,{\sc ii} $\lambda$4686 is observed in I\,Zw~18,
SBS~0335-052E and other very metal-poor star forming galaxies. 

Potentially large WR  populations are inferred in very metal-deficient 
galaxies, depending upon the exact WR wind dependence upon metallicity
 (Crowther \& Hadfield 2006). Potentially, single star rotating 
evolutionary models are unable to reproduce the observed WR distribution
in metal-poor galaxies. Close
binary evolution might represent the primary formation channel for such 
metal-poor WR  stars, unless LBV eruptions provide the dominant method of 
removing the H-rich envelope at low metallicity.

\subsection{WR stars as SNe and GRB progenitors}

The end states of massive stars have been studied from a theoretical 
perspective by Heger et al. (2003). In particular, WN and WC stars are the 
likely progenitors of (at least some) Type~Ib and Type~Ic core-collapse 
SN, respectively. This arises because, respectively hydrogen and 
hydrogen/helium are absent in such SNe (Woosley \& Bloom 2006). Direct 
empirical evidence connecting single WR stars to Type~Ib/c SN is lacking, 
for which lower mass interacting binaries represent alternative 
progenitors. 
One would need observations of $\geq 10^{4}$ WR stars in 
order to firmly establish a connection on a time frame of a few years, 
since WR lifetimes are a few 10$^{5}$~yr (Meynet \& Maeder 2005). Hadfield 
et al. (2005) identified $10^{3}$ WR stars in M83. Narrow-band optical 
surveys of a dozen other high star-forming spiral galaxies within 
$\sim$10~Mpc would likely provide the necessary statistics. However, 
ground-based surveys would be hindered by the relatively low spatial 
resolution of 20~pc per arcsec at 5~Mpc.

Nevertheless, the light curves of broad-lined Type~Ic supernovae 
-- SN~1998bw, SN~2003dh and SN~2003lw -- suggest
ejected core masses of order 10$M_{\odot}$ (Nakamura et al. 2001;  
Mazzali et al. 2003; Malesani et al. 2004). 
These agree rather well with the masses of LMC WC4
stars inferred by Crowther et al. (2002), if we additionally consider
several solar masses which remain as a compact 
(black hole) remnant. 
Tese supernovae were associated with long GRBs, namely 980425 (Galama et 
al. 1998), 030329 (Hjorth et al. 2003), and 031203 (Malesani et al. 2004), 
in support of the 
`collapsar'  model (MacFadyen \& Woosley 1999). Indeed, 
WR populations have been 
detected in the host
galaxy of GRB 980425, albeit offset from the location of the GRB by
several hundred pc (Hammer et al. 2006). Perhaps GRBs are produced
by runaway WR stars, ejected from high density star clusters (Cantiello
et al. 2007)? Such a scenario would appear to contradict
Fruchter et al. (2006), regarding the location of GRBs in their host galaxies.

The challenge faced by both single and binary evolutionary models is for
a rotating core at the point of core-collapse (Woosley \& Heger 2006). 
Single star models
indicate that stars efficiently spin-down during either: (a)
the slowly rotating RSG stage -- due to the magnetic dynamo produced by 
differential rotation between the rotating He-core and non-rotating 
H-envelope; (b) the mechanical loss of angular momentum from the core 
during the high mass-loss WR phase. Spectropolarimetry does not
favour rapid rotation for Milky Way WC stars, although some WN stars
may possess significant rotation rates. 

Typical magnetic fields of neutron stars  are of order 
10$^{12}$~G, so one would expect a field of 10$^{2}$ G  for 
their progenitor Wolf-Rayet stars. The first observational 
limits are now becoming available, namely $\leq$25 G for HD~50896 (WR6, 
St-Louis et al.  2007). If WR stars are credible progenitors of 
`magnetars', a subset of neutron stars that are highly magnetized ($\sim 
10^{15}$ G), the required WR magnetic field would be $\sim 10^{3}$ G.

Initial rapid
rotation of a single massive star may be capable of circumventing an
extended envelope via chemically homogeneous evolution (Maeder 1987) if
mixing occurs faster than the chemical gradients from nuclear fusion.
At sufficiently low metallicity,  mechanical mass-loss during the WR phase 
would be sufficiently weak to prevent loss of significant angular momentum 
permitting the necessary conditions for a GRB (Yoon \& Langer 2005).
Alternatively, close binary evolution could cause the
progenitor to spin-up due to tidal interactions or the merger of a black
hole and He core within a common envelope evolution (Podsiadlowski et al.
2004). Both single and binary scenarios may operate. At present, the 
single scenario is favoured since long-soft GRBs are 
predominantly  observed in 
host galaxies which are fainter, more irregular and more metal-deficient 
than hosts of typical core-collapse supernovae (e.g. Fruchter et al. 2006). 

Of course, the ejecta strongly interact with the circumstellar material,
probing the immediate vicinity of the GRB itself 
(van Marle, Langer \& Garc\'{i}a-Segura 2005).  This
provides information on the progenitor, for which one expects $\rho
\propto r^{-2}$ for WR winds (Eqn~\ref{continuity}). 
A metallicity-dependence of WR winds suggests that one would 
potentially expect rather different environments for the afterglows of 
long-duration GRBs, depending upon the metallicity of the host galaxy. 
Indeed, densities of the immediate environment of many GRBs suggest values 
rather lower than typical Solar metallicity WR winds (Chevalier, Li \& 
Frasson 2004). Fryer, Rockefeller \& Young (2006) estimate half of long 
GRBs apparently occur in uniform environments, favouring a post-common 
envelope binary merger model.

\section{Summary Points} 

\begin{enumerate}

\item The agreement between multi-wavelength spectroscopic observations of
WR stars and current non-LTE model atmospheres is impressive.
H-rich (core H-burning) WN stars are readily 
distinguished from classic H-deficient
(core He-burning) WN, WC and WO stars. Significant progress has 
been achieved in interior evolutionary models through the incorporation of 
rotational mixing. Contemporary (i.e. low) mass-loss 
rates together with rotational mixing permits 
many of the observed properties of WR stars to be reproduced, at least 
those close to Solar metallicity.

\item Empirical evidence and theoretical models both favour 
metallicity-dependent WR wind, providing a natural explanation to the 
WC (and WO) subtype distribution in the Milky Way and in 
external galaxies. A metallicity dependence is partially responsible for the 
WN subtype dependence, although the reduced nitrogen content in metal-poor 
galaxies also plays a role. The development of consistent 
radiatively-driven WR  winds represents an important milestone. 
The predicted  metallicity 
dependence of mass-loss rates from radiatively driven wind models agrees with 
observational estimates.

\item The apparent convergence of spectroscopic and interior
models suggests that we can re-assess the contribution of WR stars to the 
ionizing, mechanical and chemical enrichment of the ISM in young star
forming regions with respect to the previous Annual Review article on this
subject (Abbott \& Conti 1987). WR stars emit a high number of 
Lyman continuum photons due to higher inferred stellar luminosities
as a result of the inclusion of line blanketing in atmospheric models.
Inferred WR mass-loss rates have decreased due to overwhelming evidence
in favour of clumped winds. The chemical enrichment from WR  stars is
believed to be significant due to an extended WC phase, as a result of 
rotational mixing within evolutionary models.

\end{enumerate}

\section{Future Issues to be Resolved}  

\begin{enumerate} 
\item Current non-LTE models rely upon a simplified clumpy wind 
structure, plus spherical symmetry. Theoretical hydrodynamic predictions 
for the radial dependence of clumping in WR winds is  anticipated to 
represent a major area of development over the next decade together with 
two dimensional non-LTE  radiative  transfer codes. 

\item Discrepancies persist between empirical WR populations and
predictions from evolutionary models which allow for rotational mixing.
Metallicity dependent WR winds may improve consistency, together with
the incorporation of magnetic fields. 
Spatially resolved clusters rich in WR stars, such as 
Westerlund~1 (if it is genuinely co-eval), 
provide a direct means of testing evolutionary 
predictions. Observational limits on magnetic fields within
Wolf-Rayet stars are underway.

\item The role of rotation in WR stars at low metallicity is of particular
interest. At present, amongst the best means of establishing rapid 
rotation in WR
stars is by measuring departures from spherical symmetry using
spectropolarimetry. This technique has been applied to bright WR stars
in the Milky Way, but WR stars within the Magellanic Clouds are also 
within the reach of ground-based 8-10m instruments.

\item The evolutionary paths leading to WR stars and core-collapse 
SN remain uncertain, as is the role played by LBV eruptions for the most 
massive stars. Given the observational connection between Type 
Ic supernovae and  long-soft GRBs, do  GRBs result from low 
metallicity  massive stars undergoing chemically homogeneous evolution, 
massive binaries during a common envelope phase, or runaways from dense 
star clusters?

\item The presence of WR stars in large numbers within very low 
metallicity galaxies appears contrary to the expectations of single star 
evolutionary models neglecting rotational mixing. Are such stars 
exclusively produced by rapid rotation, 
close binary evolution, or via giant LBV eruptions? 

\item Most late WC stars are known to be dust formers, the 
universal
presence of binary companions is not yet established for all dusty
WC stars. In addition,
it is not clear how  dust grains form within such an extreme 
environment, the study of which merits further study.

\end{enumerate}

\section*{Acknowledgement} Thanks to G\"{o}tz Gr\"{a}fener, John Hillier,
Norbert Langer, Georges Meynet, Tony Moffat, Nathan Smith, Dany Vanbeveren
and Peredur Williams for useful comments on an early version of the
manuscript.  I wish to thank the Royal Society for providing financial
assistance through their wonderful University Research Fellowship scheme
for the past eight years.

\section*{Key Terms} 

{\bf Narrow-band imaging:} WR candidates may be identified
from narrow-band images sensitive to light from strong WR emission lines,
after subtraction of images from their adjacent continua.

\noindent {\bf P Cygni profiles}: Spectral lines showing 
blue-shifted absorption plus red-shifted 
emission. Characteristic of stellar outflows, associated with resonance 
lines of abundant ions (e.g. C\,{\sc iv} 1548-51\AA).

\noindent {\bf Non-LTE:} Solution of full rate equations is necessary due to
intense radiation field. Radiative processes dominate over collisional
processes, so Local Thermodynamic Equilibrium (LTE) is not valid.

\noindent {\bf Radiatively driven winds:} The transfer of photon momentum
in the photosphere to the stellar atmosphere through absorption by
(primarily) metal spectral lines.

\noindent {\bf Monte Carlo models:} A statistical approach to
the radiative transfer problem, using the concept of photon packets.

\noindent {\bf Clumped winds:} Radiatively driven winds are intrinsically
unstable, producing compressions and rarefactions in their outflows.

\noindent {\bf Collapsar:} Rapidly rotating WR star undergoes
core-collapse to form a black hole fed by an accretion disk, whose
rotational axis collimates the gamma ray burst jet.

\noindent {\bf Gamma Ray Burst:} Brief flash of gamma rays from
cosmological distances. Either a merger of two neutron stars 
(short burst) or a collapsar (long burst).

\noindent {\bf Magnetar:} Highly magnetized neutron star, observationally
connected with Soft Gamma Repeaters and Anomalous X-ray Pulsars.

%

\section*{Reference Annotations} 

{\bf Gr{\"a}fener \& Hamann 2005:} First solution of hydrodynamics 
within a realistic Wolf-Rayet model atmosphere.

\noindent 
{\bf Hillier 1989} Describes the extended atmospheric structure of a WC star.

\noindent 
{\bf van der Hucht 2001} Catalogue of Milky Way WR stars, including cluster 
membership, binarity, masses.

\noindent {\bf Lamers et al. 1991} Summary of key observational evidence in
favour of a late stage of evolution for WR stars.

\noindent {\bf Massey 2003} Review article on the broader 
topic of massive stars within Local Group galaxies.

\noindent {\bf Meynet \& Maeder 2005} Comparison between observed WR
populations in galaxies with evolutionary model predictions allowing for
rotational mixing.

\noindent {\bf Schaerer \& Vacca 1998} Describes the determination of WR and O
star populations within unresolved galaxies.

\noindent {\bf Williams et al. 1987} Describes various 
aspects of dust formation around WC stars.

\section*{Acronyms List} 

{\bf GRB} Gamma Ray Burst

\noindent 
{\bf LBV} 
Luminous Blue Variable

\noindent 
{\bf RSG} Red Supergiant

\noindent 
{\bf WN} Nitrogen sequence Wolf-Rayet

\noindent 
{\bf WC} Carbon sequence Wolf-Rayet

\noindent 
{\bf WO} Oxygen sequence Wolf-Rayet

\noindent 
{\bf LTE} Local Thermodynamic Equilibrium

\noindent 
{\bf IMF} Initial Mass Function

 
\section*{Side Bar} 

Luminous Blue Variables, also known as Hubble-Sandage or S~Doradus type
variables, share many characteristics of Wolf-Rayet stars. LBVs are widely
believed to be the immediate progenitors of classic WN stars. LBVs
possess powerful stellar winds, plus hydrogen depleted atmospheres,
permitting
similar analysis techniques to be used (e.g. Hillier et al. 2001).  LBVs
occupy a part of the Hertzsprung-Russell diagram adjacent to Wolf-Rayet
stars. Typical spectral morphologies vary irregularly between
A-type (at visual maximum) and B-type (at visual minimum) supergiants.
Examples include AG~Car and P~Cyg in the Milky Way, and S~Dor and R127 in
the LMC (e.g. Humphreys \& Davidson 1994). LBVs undergo occasional giant
eruptive events -- signatures of which are 
circumstellar nebulae -- most notably undergone by
$\eta$ Car during two decades in the 19th Century ($\geq10M_{\odot}$
ejected). Giant eruptions 
are believed to play a major role in the evolution of very
massive stars via the removal of their hydrogen-rich envelope (Davidson \&
Humphreys 1997; Smith et al. 2003).  The origin of such huge eruptions is
unclear, since line-driven radiation pressure is incapable of producing
such outflows.

\begin{table}
\caption{Wavelength-specific observed and synthetic 
spectral atlases (X-ray to mid-IR) of Galactic WR stars.
Predicted stellar luminosities for early-type and late-type WR
stars  within each spectral window are based upon
the averages of HD~96548 (WR40, WN8, Herald, Hillier \& Schulte-Ladbeck 2001),
HD~164270 (WR103, WC9,  Crowther, Morris \& Smith 2006a) and HD~50896 (WR6, WN4b,
Morris, Crowther \& Houck 2004), HD~37026 (BAT52, WC4, Crowther et al. 2002),
respectively. }\label{atlas}
\begin{center}
\begin{tabular}{llllll}
\toprule
$\lambda$ & Window & \multicolumn{2}{c}{$L_{\lambda}/L_{\rm bol}$} & Sp Type& Reference\\
          &        & Late & Early & \\ 
\colrule
5--25\AA\ &  X-ray  & 10$^{-7}$ &  10$^{-7}$ & WN& Skinner et al. 2002\\
          &         &           &            &  WC & Schild et al. 2004\\
$<$912\AA\ & Extreme UV & 39\% & 69\% &WN, WC& Smith, Crowther \& Norris 2002\\
              &        &    & & WN    & Hamann \& Gr\"{a}fener 2004\\
912--1200\AA\ & Far-UV &  21\% & 12\% & WN, WC&  Willis et al. 2004\\
1200--3200\AA\ & UV, Near-UV & 33\% & 16\% & WN, WC  &Willis et al. 1986\\
3200--7000\AA\ & Visual & 5\% & 2\% & WN, WC &Conti \&  Massey 1989 \\
7000--1.1$\mu$m & Far-red & 0.9\% & 0.3\%  & WN, WC &Conti, Massey \& Vreux 1990\\
                 &         &    & & WN,WC & Howarth \& Schmutz 1992\\
1--5$\mu$m &  Near-IR   &  0.4\% & 0.2\% &WN,WC &Vacca et al. 2007\\
5--30$\mu$m & Mid-IR   & 0.02\% & 0.01\%& WN & Morris et al. 2000 \\    
              &       &   & &WCd & van der Hucht et al. 1996 \\
\botrule
\end{tabular}
\end{center}
\end{table}

\begin{table}
\begin{center}
\caption{Physical and wind properties of Milky Way WR stars (LMC
in parenthesis),
adapted from  Herald, Hillier \& Schulte-Ladbeck (2001) and Hamann, 
Gr\"{a}fener \& Liermann (2006) for WN stars, plus Barniske, Hamann \& 
Gr\"{a}fener (2007), Crowther et al.  (2002, 2006a) and  references  
therein for WC stars.  Abundances are shown by mass fraction in 
percent. Mass-loss rates assume a volume filling factor of $f$=0.1.}
\label{WRtemp}
\begin{tabular}{llcccccl}
\toprule
Sp & $T_{\ast}$ 
& $\log L$ & $\dot{M}$ & $v_{\infty}$ & $\log$N(LyC) 
& $M_{v}$ & Example\\
Type & kK             & $\L_{\odot}$  & $M_{\odot}$yr$^{-1}$ & 
km\,s$^{-1}$ & ph\,s$^{-1}$ & mag             \\
\colrule
\multicolumn{8}{c}{WN stars}\\
3-w & 85& 5.34 & --5.3& 2200& 49.2 &--3.1 & WR3 \\ 
4-s & 85& 5.3  & --4.9& 1800& 49.2 &--4.0 & WR6 \\ 
5-w & 60& 5.2  & --5.2& 1500& 49.0 &--4.0 & WR61   \\ 
6-s & 70& 5.2  & --4.8& 1800& 49.1 &--4.1 & WR134   \\ 
7   & 50& 5.54 & --4.8&1300 & 49.4 &--5.4 & WR84    \\ 
8   & 45& 5.38 & --4.7&1000 &49.1  &--5.5 & WR40    \\ 
9   & 32& 5.7  & --4.8& 700 &48.9  &--6.7 & WR105      \\ 
\multicolumn{8}{c}{WNha stars}\\
6ha & 45& 6.18 & --5.0& 2500 &49.9 &--6.8 & WR24    \\ 
9ha & 35& 5.86 & --4.8&1300  &49.4 &--7.1 & WR108  \\ 
\multicolumn{8}{c}{WC and WO stars}\\
(WO) & (150)& (5.22) &  (--5.0) & (4100) & (49.0) & (--2.8) &(BAT123)\\
(4) &(90)&  (5.54)& (--4.6) & (2750) & (49.4) &(--4.5)& (BAT52)     \\ 
5  & 85 & 5.1 & --4.9 & 2200 & 48.9 &--3.6& WR111      \\ 
6  & 80 & 5.06 & --4.9 & 2200 & 48.9 &--3.6& WR154       \\ 
7  & 75 & 5.34 & --4.7 & 2200 & 49.1 &--4.5& WR90        \\ 
8  & 65 & 5.14 & --5.0 & 1700 & 49.0 &--4.0& WR135     \\ 
9  & 50 & 4.94 & --5.0 & 1200 & 48.6 &--4.6& WR103      \\ 
\botrule
\end{tabular}
\end{center}
\end{table}

\clearpage

\end{document}